\documentclass[10pt,twocolumn,journal]{IEEEtran}
\usepackage{setspace}
\usepackage{clrscode}
\usepackage{pict2e}

\usepackage{amsmath}
\usepackage{amssymb}
\usepackage[dvips]{graphicx}
\usepackage{epsfig}
\usepackage{algorithm}
\usepackage{algorithmic}
\usepackage{caption}
\usepackage{amsthm}
\usepackage{subcaption}
\usepackage{mathrsfs}

\usepackage[ps2pdf,
bookmarks=false,
bookmarksnumbered=false, 
bookmarksopen=false, 
colorlinks=false]{}

\newcommand{\rme}{{\mathrm{e}}}

\newcommand{\E}{\mathbb{E}}

\newcommand{\mH}{\mathcal{H}}
\newcommand{\hH}{\hat{\mathcal{H}}}

\makeatletter

\newcommand{\Rmnum}[1]{\expandafter\@slowromancap\romannumeral #1@}
\makeatother

\newcommand{\argmax}{\operatornamewithlimits{arg\,max}}

\newcommand{\avg}{\text{avg}}

\newcommand{\nid}{\rm{d}}
\newcommand{\f}{\rm{f}}

\newcommand{\pk}{\rm{pk}}

\newtheorem{Lem1}{Theorem}

\newtheorem{Rem}{Remark}

\newtheorem{Prop}{Proposition}

\begin{document}

%
\title{Spectral and Energy Efficiency in Cognitive Radio Systems with Unslotted Primary Users and Sensing Uncertainty}

\author{
\IEEEauthorblockN{Gozde Ozcan, M. Cenk Gursoy, and Jian Tang}
\thanks{Gozde Ozcan, M. Cenk Gursoy and Jian Tang are with the Department of Electrical
Engineering and Computer Science, Syracuse University, Syracuse, NY, 13244
(e-mail: gozcan@syr.edu, mcgursoy@syr.edu, jtang02@syr.edu).}}

\maketitle

\begin{abstract}\let\thefootnote\relax\footnote{The material in this paper was presented in part at the IEEE International Conference on Communications (ICC), London, UK, June 2015.}
\let\thefootnote\relax\footnote{This research is supported by NSF grants CNS-1443966 and ECCS-1443994. }
This paper studies energy efficiency (EE) and average throughput maximization for cognitive radio systems in the presence of unslotted primary users. It is assumed that primary user activity follows an ON-OFF alternating renewal process. Secondary users first sense the channel possibly with errors in the form of miss detections and false alarms, and then start the data transmission only if no primary user activity is detected. The secondary user transmission is subject to constraints on collision duration ratio, which is defined as the ratio of average collision duration to transmission duration. In this setting, the optimal power control policy which maximizes the EE of the secondary users or maximizes the average throughput while satisfying a minimum required EE under average/peak transmit power and average interference power constraints are derived. Subsequently, low-complexity algorithms for jointly determining the optimal power level and frame duration are proposed. The impact of probabilities of detection and false alarm, transmit and interference power constraints on the EE, average throughput of the secondary users, optimal transmission power, and the collisions with primary user transmissions are evaluated. In addition, some important properties of the collision duration ratio are investigated. The tradeoff between the EE and average throughput under imperfect sensing decisions and different primary user traffic are further analyzed.
\end{abstract}

\thispagestyle{empty}

\section{Introduction}
Cognitive radio is a promising innovative technology, leading to more efficient spectrum management and utilization. In cognitive radio systems, unlicensed users (i.e., cognitive or secondary users) are allowed to either continuously share the spectrum licensed with legacy users (i.e., primary users) without causing any significant interference, or periodically monitor the primary user activity via spectrum sensing and then perform transmissions according to sensing decisions.

\subsection{Motivation}
Increasing global energy demand, consequent environmental concerns in terms of high levels of greenhouse gas emissions, high energy prices, and operating costs currently have triggered extensive research efforts in energy efficient communication systems. Optimal and efficient use of energy resources is paramount importance for cognitive radio systems in order to effectively utilize limited transmission power of battery-powered cognitive radios and support additional signal processing requirements such as spectrum sensing. Many existing works have focused on the design of optimal resource allocation schemes to maximize the spectrum efficiency (SE) and energy efficiency (EE) of the cognitive users. In particular, the authors in \cite{stotas} determined the optimal power control and sensing duration to maximize the ergodic capacity of cognitive radio systems operating in multiple narrowband channels under two different transmission schemes, namely sensing-based spectrum sharing and opportunistic spectrum access. The authors in \cite{akin} characterized the effective capacity of secondary users and the corresponding optimal power control policy in the presence of sensing errors.

The work in \cite{hu} mainly focused on the design of the optimal sensing duration and sensing decision threshold to maximize the weighted sum of the EE and SE. The authors in \cite{zhang} analyzed the optimal sensing duration that maximizes the EE of secondary user subject to a constraint on the detection probability. In \cite{xiong}, the optimal subcarrier assignment and power allocation were proposed to maximize the worst case EE, (i.e., by considering the secondary user with the lowest EE) or to maximize the average EE of secondary users in an OFDM-based cognitive radio network. The work in \cite{park} studied the optimal power control scheme that maximizes the sum of EEs of the cognitive femto users for 5G communications. The authors in \cite{ramamonjison} developed energy-efficient power control algorithms for secondary users in a two-tier cellular network. In these works, it is assumed that primary users transmit in a time-slotted fashion, i.e., the activity of the primary users (e.g., active or inactive) remains the same during the entire frame duration.

In practice, primary and secondary user transmissions may not necessarily be synchronized. For instance, the primary user traffic can be bursty and may change its status during the transmission phase of the secondary users. In such cases, the assumption of time-slotted primary user transmission adopted in most studies (as also seen in the above-mentioned works) does no longer hold. In unslotted scenarios, it is assumed that ON-OFF periods of the primary user transmissions are random variables, following certain specific distributions. Exponential distribution is a commonly used model (see e.g., \cite{pei} -- \cite{zarrini}). In particular, the authors in \cite{pei} determined the optimal frame duration that maximizes the throughput of the secondary users with perfect sensing decisions under collision constraints, assuming that the primary user activity changes only once within each frame. By adopting the same assumptions for the primary user activity as in the previous work, the authors in \cite{wang} mainly focused on the throughput of secondary users operating in the presence of multiple primary users with imperfect channel sensing results. In the same setting, the work in \cite{zarrini} mainly analyzed the optimal frame duration that maximizes the secondary user throughput. In \cite{guerrini}, the exact secondary user throughput was determined and joint optimization of the sensing duration and frame period in the presence of sensing errors was performed by assuming that the primary user changes its status multiple times. The authors in \cite{macdonald} -- \cite{dhakal} studied the impact of primary user activity on the sensing performance. The works in \cite{tang}, \cite{pradhan} analyzed the sensing-throughput tradeoff for a secondary user in the presence of random arrivals and departures of the primary user and multiple transitions of primary user activity during sensing duration.

\subsection{Main Contributions}
The recent work in \cite{zhang2} analyzed general EE-SE relation for overlay, underlay and interweave cognitive radio systems. In particular, the authors introduced a general EE optimization problem and derived closed-form EE expressions for these systems. However, the authors did not consider collision constraints in the optimization problem, the transmission power was not instantaneously adapted according to channel conditions and the perfect sensing was assumed for interweave cognitive radio systems. In practice, cognitive radio systems, which employ spectrum sensing mechanisms to learn the channel occupancy by primary users, generally operate under sensing uncertainty arising due to multipath fading, shadowing and hidden node problem. Such kind of events can be incorporated into sensing uncertainty \cite{sharma}, \cite{kaushik}. Therefore, motivated mainly by the fact that the optimal power control policies that maximize the EE or maximize the SE under constraints on EE (and hence address the tradeoff between EE and SE) have not been derived in the presence of unslotted primary users and imperfect sensing results, we have the following key contributions in this paper:
\begin{itemize}
\item We derive, in closed-form, the optimal power control policy that maximizes the EE of the secondary users operating with unslotted primary users subject to peak/average transmit power, average interference power and collision constraints in the presence of sensing errors. Hence, the power level has been adapted instantaneously according to the channel power gains of both the transmission link between the secondary transmitter and the secondary receiver and the interference link between the secondary transmitter and the primary receiver. We do not impose any limitations on the number of transitions of the primary user activity unlike the studies in \cite{pei} -- \cite{zarrini} where the primary user activity changes only once. We assume that the primary user can change its status between ON and OFF states multiple times.

\item In order to consider the EE and SE requirements of the secondary users jointly, we obtain the optimal power control scheme that maximizes the average througput of the secondary users while satisfying the minimum required EE in the presence of unslotted primary users.

\item We propose low-complexity algorithms for jointly finding the optimal power control policy and frame duration.

\item We analyze several important properties of the collision duration ratio and relations among sensing performance, secondary user throughput, EE, optimal frame duration and the resulting collisions with the primary user.
\end{itemize}

The rest of the paper is organized as follows. Section \ref{sec:system_model} introduces the primary user activity model, opportunistic spectrum access scheme and collision constraints. In Sections \ref{sec:EE_power} and \ref{sec:SE_power}, optimal power control schemes that maximize the EE of the secondary users and the average throughput under a minimum EE constraint are derived, respectively. The algorithms for jointly determining the optimal power control and frame duration are also developed. Numerical results are provided and discussed in Section \ref{sec:num_results} before giving the main concluding remarks in Section \ref{sec:conclusion}.

\section{System Model} \label{sec:system_model}
In this paper, we consider a cognitive radio system consisting of a pair of primary transmitter and receiver, and a pair of secondary transmitter and receiver. Secondary users opportunistically access the channel licensed to the primary users. In the following subsections, we describe the primary user activity model, opportunistic spectrum access policy of the secondary users, and the formulation of the collision constraint imposed for the protection of the primary users.
\subsection{Primary User Activity Model}
\begin{figure*}[h]
\centering
\includegraphics[width=0.8\textwidth]{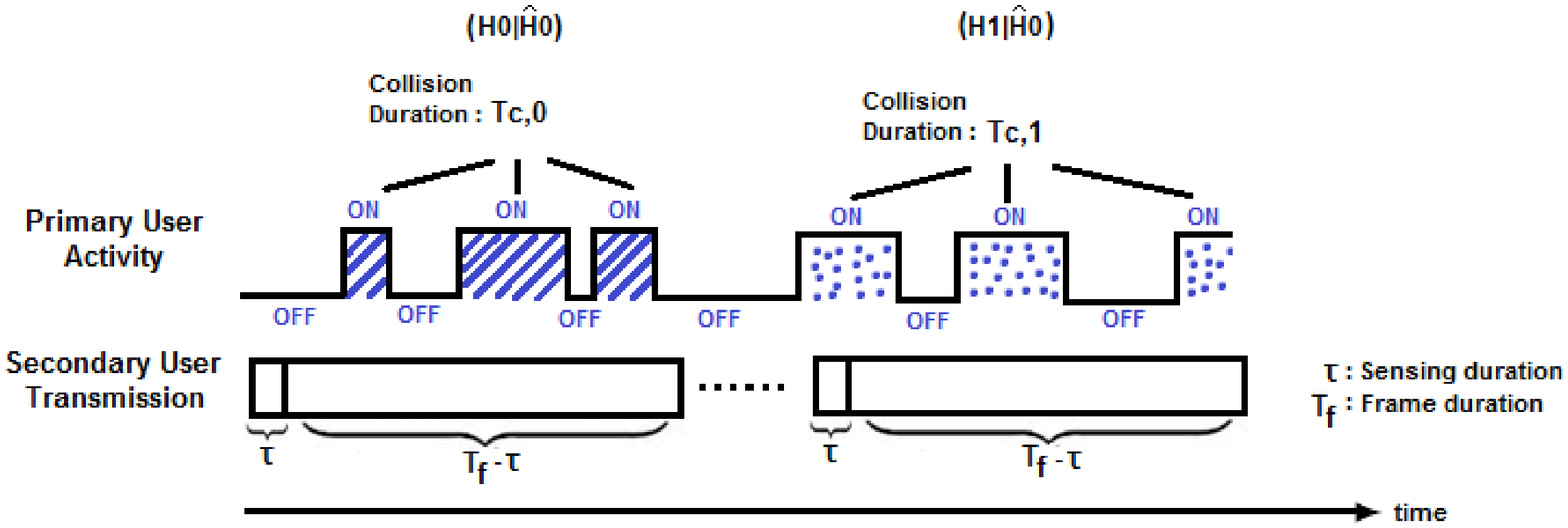}
\caption{Frame structure of the primary and secondary users.}
\label{fig:frame_PU_SU}
\vspace{-0.7cm}
\end{figure*}

Differing from the majority of the studies (which assume that the primary users adopt a time-slotted transmission scheme), we consider a continuous, i.e., unslotted transmission structure as shown in Fig. \ref{fig:frame_PU_SU} at the top of next page.

We assume that the primary user activity follows a semi-Markov process with ON and OFF states. We have adopted exponential distribution model for the primary user traffic due to its popularity and existence in real systems. In particular, the recent works \cite{riihijarvi} -- \cite{yin} have confirmed the exponential distribution for time-domain utilization of certain licensed channels through experimental simulation results. Also, the recent measurement study \cite{willkomm} has shown the exponential distribution of call arrival times in CDMA-based systems. Also, the exponentially distributed traffic model for the primary user is common assumption for cognitive radio systems in the literature. The parameters of the exponentially distributed primary user traffic can be found using blind and non-blind algorithms based on maximum likelihood estimation and adaptive sampling techniques as proposed in \cite{gabran}. In this model, the ON state indicates that the primary user is transmitting while the OFF state represents that the channel is not occupied by the primary user. Such a process is also known as an alternating renewal process. The durations of ON and OFF periods are independent of each other and are exponentially distributed with means $\lambda_0$ and $\lambda_1$, respectively, and therefore have probability density functions
\begin{align}
f_{\text{ON}}(t)=\frac{1}{\lambda_0} \rme^{-\frac{t}{\lambda_0}}, \text{ and }
f_{\text{OFF}}(t)=\frac{1}{\lambda_1} \rme^{-\frac{t}{\lambda_1}}.
\end{align}
Hence, the prior probabilities of channel being vacant or occupied by the primary user can be expressed, respectively, as
\begin{align}\label{eq:prior_probs}
\Pr\{\mH_0\}=\frac{\lambda_0}{\lambda_0+\lambda_1}, \hspace{0.5cm} \Pr\{\mH_1\}&=\frac{\lambda_1}{\lambda_0+\lambda_1}.
\end{align}

\subsection{Opportunistic Spectrum Access by the Secondary Users}
Secondary users employ frames of duration $T_{\f}$. In the initial duration of $\tau$ seconds, secondary users perform channel sensing and monitor the primary user activity. Subsequently, data transmission starts in the remaining frame duration of $T_{\f}-\tau$ seconds only if the primary user activity is not detected, the event of which is denoted by $\hH_0$. In our analysis, we consider that sensing duration is much shorter than the mean duration of ON period of the primary user traffic, therefore it is reasonable to assume that the primary user activity is constant during the sensing duration. Spectrum sensing is modeled as a simple binary hypothesis testing problem with two hypotheses $\mH_0$ and $\mH_1$ corresponding to the absence and presence of the primary user signal, respectively. Many spectrum sensing methods have been proposed \cite{axell}, and the corresponding sensing performance is characterized by two parameters, namely the probabilities of detection and false alarm, which are defined as
\begin{align}
P_{\nid}&= \Pr\{\hH_1|\mH_1\}, \hspace{0.5cm} P_{\f}=\Pr\{\hH_1|\mH_0\},
\end{align}
where $\hH_1$ denotes the event that the primary user activity is detected. We note that any sensing method can be employed in the rest of the analysis since the results depend on the sensing performance only through the probabilities of detection and false alarm, and the sensing duration.

\subsection{Collision Constraints}
We first describe the secondary users' collisions with the primary users, which can lead to considerable performance degradation in the primary user communication. Subsequently, we impose a constraint on the ratio of the average collision duration to the transmission duration in order to protect the primary users. Depending on the true nature of the primary user activity at the beginning of the frame, collisions between the primary and secondary users can occur in the following two cases:
\begin{itemize}
\item \emph{Case $1$}: The channel is not occupied by the primary user and is correctly detected as idle at the beginning of the frame. Even if the primary user is not actually transmitting initially, it is possible for the primary user to start data transmission at any time during the current frame, which results in a collision event. By conditioning on the correct detection of the initial absence of the primary user, the ratio of the average collision duration to data transmission duration, which is called the collision duration ratio, can be expressed as
\begin{align}
\mathscr{P}_{c,0}=\frac{\E\{T_{c | \mH_0, \hH_0}\}}{T_{\f}-\tau},
\end{align}
where $\E\{\cdot\}$ denotes the expectation, and $T_{c | \mH_0,\hH_0}$ is a random variable representing the collision duration between the secondary and primary users given that the primary user is inactive initially at the beginning of the frame (event $\mH_0$) and the sensing decision is idle (event $\hH_0$). It is assumed that the primary user is in the OFF state at first and taking into account the possible multiple transitions between ON and OFF states. In this setting, the recursive expression of $T_{c | \mH_0,\hH_0}$ is written as
\begin{align} 
\small
\begin{split}\label{eq:recursive_collision}
\hspace{-0.4cm}T_{c | \mH_0,\hH_0} (t|X,\!Y) \!=\!\!\begin{cases}0 &t \leq X \\
 T_{\f}-\tau-X  \hspace{1cm}&X \! \leq \! t \! \leq \! X\!+\!Y\\
Y\!+\!T_{c | \mH_0,\hH_0} (t\!-\!X\!-\!Y) &X+Y \leq t,
 \end{cases}
\end{split}
\normalsize
\end{align}
where $X$ denotes the first OFF state, which is exponentially distributed with mean $\lambda_1$ and
and $Y$ represents the first ON state, which is exponentially distributed with mean $\lambda_0$. Using (\ref{eq:recursive_collision}), $\E\{T_{c | \mH_0, \hH_0}\}$ is calculated in the following:
\begin{align}
\E\{T_{c | \mH_0, \hH_0}\} = \int \int_{xy} T_{c | \mH_0,\hH_0} (t|X,Y) f_{XY}(x,y)dxdy.
\end{align}
Then, the closed form expression for $\E\{T_{c |\mH_0,\hH_0}\}$ can be found by using Laplace transform and following the same steps as in \cite[Theorem 2]{jiang}. Hence, $P_{c,0}$ is given by
\begin{align}
\mathscr{P}_{c,0}=\Pr\{\mH_1\}-\frac{\lambda_0 \Pr\{\mH_1\}^2}{T_{\f}-\tau}\bigg(1-\rme^{-\frac{T_{\f}-\tau}{\lambda_0 \Pr\{\mH_1\}}}\bigg).
\end{align}
\item \emph{Case $2$}: The primary user is actually present in the channel at the beginning of the frame, however the secondary user miss-detects the primary user activity, resulting in a collision right away due to sensing error. Multiple collisions can also occur if the primary user turns OFF and then back ON in a single frame once or multiple times. Similar to the first case, by conditioning on the miss detection event, the collision duration ratio can be found as
\begin{align}
\mathscr{P}_{c,1}&=\frac{\E\{T_{c |\mH_1,\hH_0}\}}{T_{\f}-\tau}
\\
&=\Pr\{\mH_1\}\!+\!\frac{\lambda_1 \Pr\{\mH_0\}^2}{T_{\f}-\tau}\bigg(1\!-\!\rme^{-\frac{T_{\f}-\tau}{\lambda_0 \Pr\{\mH_1\}}}\bigg)
\end{align}
where  $T_{c | \mH_1, \hH_0}$ is a random variable describing the collision duration between the secondary and primary users given that the primary user is active at the beginning of the frame but sensing decision is incorrectly an idle channel.
\end{itemize}
Based on the above two cases, the collision duration ratio averaged over the true nature of the primary user activity given the idle sensing decision $\hH_0$ can be expressed as
\begin{align}
\mathscr{P}_c= \Pr\{\mH_0|\hH_0\}\mathscr{P}_{c,0}+\Pr\{\mH_1|\hH_0\}\mathscr{P}_{c,1}
\end{align}
where $\Pr\{\mH_0|\hH_0\}$ and $\Pr\{\mH_1|\hH_0\}$ denote the conditional probabilities of the primary user being active or inactive given the idle sensing decision, respectively, which can be written in terms of $P_{\nid}$ and $P_{\f}$ as
\begin{align} \label{eq:PH0_hH0_exp}
\Pr\{\mH_0|\hH_0\}&=\frac{\Pr\{\mH_0\}(1-P_{\f})}{\Pr\{\mH_0\}(1-P_{\f})+\Pr\{\mH_1\}(1-P_{\nid})},
\end{align}
\begin{align} 
\label{eq:PH1_hH0_exp}
\Pr\{\mH_1|\hH_0\}&=\frac{\Pr\{\mH_1\}(1-P_{\nid})}{\Pr\{\mH_0\}(1-P_{\f})+\Pr\{\mH_1\}(1-P_{\nid})}.
\end{align}
In the following, we provide two key properties of $\mathscr{P}_c$.
\begin{Prop} \label{prop:1} The average collision duration ratio $\mathscr{P}_c$ under idle sensing decision has the following properties:
\begin{itemize}
\item It is an increasing function of the frame duration $T_{\f}$ for $P_{\f}<P_{\nid}$ and a decreasing function for $P_{\f}>P_{\nid}$.
\item It takes values between $\Pr\{\mH_1|\hH_0\}$ and $\Pr\{\mH_1\}$.
\end{itemize}
\end{Prop}
\emph{Proof:} See Appendix \ref{appendix1}.

\begin{figure}[htb]
\centering
\includegraphics[width=0.5\textwidth]{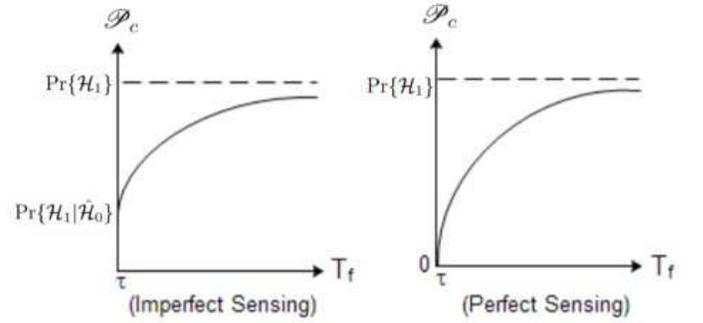}
\caption{Average collision duration vs. frame duration $T_{\f}$ in the cases of imperfect sensing and perfect sensing.}
\label{fig:Pc_Tf_illustrate}
\end{figure}

In Fig. \ref{fig:Pc_Tf_illustrate}, we illustrate $\mathscr{P}_c$ as a function of the frame duration $T_{\f}$ when $P_{\f}<P_{\nid}$, i.e., correct detection probability is greater than the false alarm probability. Note that this is generally the desired case in practice in which the probability of detection is expected to be greater than $0.5$ and the probability of false alarm be less than $0.5$ for reliable sensing performance. In the figure, both imperfect sensing and perfect sensing are considered. For the case of imperfect sensing, $\mathscr{P}_c$ takes values between $\Pr\{\mH_1|\hH_0\}$ and $\Pr\{\mH_1\}$. For perfect sensing, $\mathscr{P}_c$ is first $0$ since $\Pr\{\mH_1|\hH_0\}=0$, which corresponds to no collision event initially, as expected, and then $\mathscr{P}_c$ starts to increase with increasing $T_{\f}$ as it becomes more likely that the primary user initiates a transmission and secondary users collide with the primary users.

\section{Energy-Efficient Optimal Power Control and Frame Duration} \label{sec:EE_power}
\subsection{Average Transmit Power and Average Interference Power Constraints}
In this subsection, we determine the optimal power control policy and frame duration that maximize the EE of the secondary users in the presence of sensing uncertainty and unslotted primary users. We consider average transmit power and average interference power constraints. The latter constraint is imposed by the secondary transmitter to maintain a long-term power budget and hence long battery life by limiting the average transmit power by $P_{\avg}$, which is the maximum average transmit power limit. The former constraint is imposed to satisfy the long-term QoS requirements of the primary users by limiting the average interference power by $Q_{\avg}$, which represents the maximum average received interference power limit at the primary receiver. Regulatory bodies (e.g., Federal Communication Commissions (FCC)) sets an interference temperature limit, $Q_{\avg}$ which provides the maximum amount of tolerable interference at the primary receiver for a given frequency band at a particular location. In this setting, the optimization problem can be formulated as
\begin{align}
\label{eq:EE_max_Pavg_Qavg}
&\hspace{-2cm}\max_{
\substack{T_{\f}, P(g,h)}} \eta_{EE}=\frac{R_{\avg}}{\big(\frac{T_{\f}-\tau}{T_{\f}}\big)P(\hH_0)\E_{g,h}\{P(g,h)\}+P_{c_r}} \\  \label{eq:coll_constraint}
\text{subject to}\hspace{0.4cm}&\mathscr{P}_c \leq \mathscr{P}_{c,max} \\  \label{eq:avg_transmit_power}
&\Big(\frac{T_{\f}-\tau}{T_{\f}}\Big)\Pr\{\hH_0\}\E_{g,h}\big\{P(g,h)\big\} \le P_{\avg} 
\end{align}
\begin{align} \label{eq:avg_inter_power}
&\Big(\frac{T_{\f}-\tau}{T_{\f}}\Big)\mathscr{P}_c\Pr\{\hH_0\}\E_{g,h}\big\{P(g,h)|g|^2\big\} \le Q_{\avg} \\
& P(g,h)  \geq 0 \\
& T_{\f}  \geq \tau,
\end{align}
where the EE in the objective function is defined as the ratio of average throughput of the secondary users to the total power consumption, including average transmission power and circuit power, denoted by $P_{c_r}$.
Above, $P(g,h)$ denotes the instantaneous transmission power as a function of the channel fading coefficient $g$ of the interference link between the secondary transmitter and the primary receiver, and the channel fading coefficient $h$ of the transmission link between the secondary transmitter and the secondary receiver. The average transmission rate expression, $R_{\avg}$ is given in (\ref{eq:Ravg}),
\begin{figure*}
\begin{equation}
\small
\begin{split} \label{eq:Ravg}
R_{\avg}=\Big(\frac{T_{\f}-\tau}{T_{\f}}\Big)\E_{g,h}\Big\{&\Pr\{\mH_0\}(1-P_{\f})\Big[\log_2\Big(1+\frac{P(g,h) |h|^2}{N_0}\Big)(1-\mathscr{P}_{c,0})+\log_2\Big(1+\frac{P(g,h) |h|^2}{N_0+\sigma_s^2}\Big)\mathscr{P}_{c,0}\Big]\\   +&\Pr\{\mH_1\}(1-P_{\nid})\Big[\log_2\Big(1+\frac{P(g,h) |h|^2}{N_0}\Big)(1-\mathscr{P}_{c,1})+\log_2\Big(1+\frac{P(g,h) |h|^2}{N_0+\sigma_s^2}\Big)\mathscr{P}_{c,1}\Big]\!\Big\},
\end{split}
\normalsize
\end{equation}
\hrule
\end{figure*}
\normalsize
where $N_0$ and $\sigma_s^2$ represent the variances of the additive Gaussian noise and primary user's received faded signal, respectively. It is assumed that the secondary transmitter has perfect channel side information (CSI), i.e., perfectly knows the values of $g$ and $h$. While the assumption of perfect CSI is idealistic, channel knowledge can be obtained rather accurately if the mobility in the environment and channel variations are relatively slow. More specifically, secondary transmitter can acquire channel knowledge once the secondary receiver learns the channel and sends this information via an error-free feedback link. Also, the knowledge of the interference link, $g$, can be obtained through direct feedback from the primary receiver \cite{alouini}, indirect feedback from a third party such as a band manager \cite{peha} or by periodically sensing the pilot symbols sent by the primary receiver under the assumption of channel reciprocity \cite{zhao}.

In (\ref{eq:coll_constraint}), $\mathscr{P}_{c,max}$ denotes the maximum tolerable collision duration ratio, which needs to be greater than $P(\mH_1 | \hH_0)$ based on Proposition \ref{prop:1} because, otherwise, the constraint cannot be satisfied. Since $\mathscr{P}_c$ is an increasing function of $T_{\f}$ when $P_{\f}<P_{\nid}$, the collision constraint in (\ref{eq:coll_constraint}) provides an upper bound on the frame duration $T_{\f}$ as follows:
\begin{align}
T_{\f} \leq \mathscr{P}_c^{-1}(\mathscr{P}_{c,max}).
\end{align}
Above, $\mathscr{P}_c^{-1}(.)$ is the inverse function of $\mathscr{P}_c$.

As the frame duration increases, the secondary users have more time for data transmission, which leads to higher throughput, consequently higher EE. On the other hand, the primary user is more likely to become active with increasing transmission duration. In this case, the secondary users may collide with the primary transmission more frequently, which reduces the throughput, and hence EE. Therefore, there indeed exists an optimal frame duration that achieves the best tradeoff between the EE of the secondary users and collisions with the primary users. It can be easily verified that the EE is not a concave function of the frame duration $T_{\f}$ since the second derivative of the EE with respect $T_{\f}$ is less than, greater than or equal to zero depending on the values of the sensing parameters and prior probabilities of primary user being active and idle. However, the optimal frame duration which maximizes the EE can easily be obtained using a one-dimensional exhaustive search within the interval $(\tau, \mathscr{P}_c^{-1}(\mathscr{P}_{c,max})]$. For a given frame duration, we derive the optimal power control policy in the following result.

\begin{Lem1}\label{teo:1} The optimal power control that maximizes the EE of the secondary users operating subject to the average transmit power constraint in (\ref{eq:avg_transmit_power}) and average interference power constraint in (\ref{eq:avg_inter_power}) in the presence of sensing errors and unslotted primary users is given by
\begin{align} \label{eq:opt_power}
&\hspace{1.8cm}P_{\text{opt}}(g,h) =\bigg[\frac{A_0+\sqrt{\Delta_0}}{2}\bigg]^{+} \\
\text{where } &A_0=\frac{\log_2(e)}{(\alpha+\lambda)+\nu\mathscr{P}_c|g|^2}-\frac{2N_0+\sigma_s^2}{|h|^2} \\
&\hspace{-1cm}\Delta_0\!=\!A_0^2\!-\!\frac{4}{|h|^2}\bigg(\!\frac{N_0(N_0\!+\!\sigma_s^2)}{|h|^2}\!-\frac{\log_2(e)(N_0+(1-\mathscr{P}_c) \sigma_s^2)}{(\alpha+\lambda)+\nu\mathscr{P}_c|g|^2}\!\bigg).
\end{align}
Above, $(x)^+=\max\{0,x\}$ and $\alpha$ is a nonnegative parameter. Morever, $\lambda$ and $\nu$ are the Lagrange multipliers which can be jointly obtained by inserting the above optimal power control into the constraints in (\ref{eq:avg_transmit_power}) and (\ref{eq:avg_inter_power}), respectively.
\end{Lem1}

\emph{Proof:} See Appendix \ref{appendix2}.

\begin{table}
\caption{} \label{table:algorithm1}
\begin{algorithm}[H]
    \caption{The optimal power control and frame duration algorithm that maximizes the EE of the secondary users under the average transmit power, average interference power, and collision constraints}
    \begin{algorithmic}[1]
      \STATE Initialize $P_{\nid}=\mathscr{P}_{\nid,init}$, $P_{\f}=\mathscr{P}_{\f,init}$, $\epsilon > 0$, $\delta > 0$, $t > 0$, $\alpha^{(0)}=\alpha_{\text{init}}$,  $\lambda^{(0)}=\lambda_{\text{init}}$, $\nu^{(0)}=\nu_{\text{init}}, \mathscr{P}_{c,max}=\mathscr{P}_{c,max,\text{init}}$
     \IF{$\mathscr{P}_{c,max} < \Pr\{\mH_1|\hH_0\}$}
      \STATE $T_{\f,\text{opt}}=0, P_{\text{opt}}(g,h)=0$
     \ELSE
\STATE $k \leftarrow 0$
\REPEAT
      \STATE $n \leftarrow 0$
\REPEAT
\STATE calculate $P_{\text{opt}}(g,h)$ using (\ref{eq:opt_power});
\STATE update $\lambda$ and $\nu$ using subgradient method as follows:
	 \STATE {\footnotesize $\lambda^{(n+1)}\!=\!\Big(\lambda^{(n)}\!-t\Big(P_{\avg}-\Big(\!\frac{T_{\f}-\tau}{T_{\f}}\!\Big)\Pr\{\hH_0\}\E_{g,h}\big\{P_{\text{opt}}(g,h)\big\}\Big)\Big)^{\!+}$ }
	\STATE {\footnotesize $\nu^{(n+1)}\!\!=\!\!\Big(\!\nu^{(n)}\!-t\Big(\!Q_{\avg}-\!\Big(\!\frac{T_{\f}\!-\!\tau}{T_{\f}}\!\Big)\!\mathscr{P}_c\!\Pr\{\!\hH_0\!\}\E_{g,h}\big\{\!P_{\text{opt}}(g,\!h)|g|^2\!\big\}\!\Big)\!\Big)^{\!+}$}
\STATE $n \leftarrow n+1$
       \UNTIL{$\Big|\lambda^{(n)}(P_{\avg}-\Big(\frac{T_{\f}-\tau}{T_{\f}}\Big)\Pr\{\hH_0\}\E_{g,h}\big\{P_{\text{opt}}(g,h)\big\}\Big)\Big| \hspace{-0.1cm}\le \delta$ and $\Big|\nu^{(n)}\Big(Q_{\avg}-\Big(\frac{T_{\f}-\tau}{T_{\f}}\Big)\mathscr{P}_c\Pr\{\hH_0\}\E_{g,h}\big\{P_{\text{opt}}(g,h)|g|^2\big\} \!\Big)\Big| \le \delta$}
\STATE $\alpha^{(k+1)}=\frac{R_{\avg}}{\big(\frac{T_{\f}-\tau}{T_{\f}}\big)\Pr\{\hH_0\}\E_{g,h}\{P_{\text{opt}}(g,h)\}+P_{c_r}}$
\STATE $k \leftarrow k+1$
       \UNTIL{$|F(\alpha^{(k)})| \le \epsilon$}
\STATE $\eta_{EE}= \alpha^{(k)}$
\STATE $T_{\f,opt}=\argmax \eta_{EE}$ by bisection search
\STATE $P^*_{\text{opt}}(g,h)=[P_{\text{opt}}(g,h)]_{T_{\f}=T_{\f,opt}}$
 \ENDIF
    \end{algorithmic}
  \end{algorithm}
\vspace{-0.9cm}
\end{table}

The values of $\lambda$ and $\nu$ can be obtained numerically via the projected subgradient method. In this method, $\lambda$ and $\nu$ are updated iteratively in the direction of a negative subgradient of the Lagrangian function $\mathcal{L}(P(g,h),\lambda,\nu,\alpha)$ (given in (\ref{eq:lagrangian_function}) in Appendix \ref{appendix2}) until convergence as follows:

\begin{align} 
\small
\begin{split} \label{eq:subgradient}
&\lambda^{(n+1)}=\Big(\lambda^{(n)}-t\Big(P_{\avg}-\Big(\frac{T_{\f}-\tau}{T_{\f}}\Big)\Pr\{\hH_0\}\E_{g,h}\big\{P(g,h)\big\}\Big)\Big)^+ 
\end{split}
\normalsize
\end{align}

\begin{align} 
\small
\begin{split}  \label{eq:subgradient_Qavg}
\nu^{(n+1)}\!=\!\!\Big(\!\nu^{(n)}\!-\!t\Big(\!Q_{\avg}\!-\!\!\Big(\frac{T_{\f}\!-\!\tau}{T_{\f}}\Big)\mathscr{P}_c\Pr\{\hH_0\}\E_{g,h}\big\{\!P(g,h)|g|^2\!\big\} \!\Big)\!\Big)^{\!+},
\end{split}
\normalsize
\end{align}
where $n$ is the iteration index and $t$ is the step size. For a constant $t$, $\lambda$ and $\nu $ are shown to converge to the optimal values within a small range \cite{boyd2}.

In Table \ref{table:algorithm1}, we provide our low-complexity algorithm for jointly finding the optimal power control policy and frame duration, which maximize the EE of the secondary users in the presence of unslotted primary users and imperfect sensing decisions. In the table, for a given value of $\alpha$ and frame duration $T_{\f}$, the optimal power control is obtained when $F(\alpha) \leq \epsilon$ is satisfied, where $F(\alpha)$ is defined in (\ref{eq:F_alpha}) in Appendix \ref{appendix2} and $\alpha$ is a nonnegative parameter. The solution is optimal if $F(\alpha)=0$, otherwise $\epsilon$-optimal solution is obtained.

The proposed power control algorithm consists of two nested loops. In the outer loop, Dinkelbach's method iteratively solves the energy efficiency maximization problem by solving a sequence of parameterized concave problems. It is shown that Dinkelbach's method has a super-linear convergence rate \cite{schaible}, and hence the sequence converges to an optimal solution in a small number of iterations. In the inner loop, Lagrange multipliers are updated using the subgradient method, which involves the computation of subgradient and simple projection operations. The subgradient method is widely used to find Lagrange multipliers due to its simplicity, easy implementation, the speed for computing a direction, and the global convergence property \cite{zhao}. In addition, the optimal frame duration is obtained by bisection search, which is the simplest root finding method. In particular, the bisection method halves the search interval at each iteration and its time complexity is logarithmic. Hence, the proposed algorithm is computationally efficient. 

\begin{Rem}
The optimal power control policy in (\ref{eq:opt_power}) is a decreasing function of average collision duration ratio, $\mathscr{P}_c$. In particular, when the secondary users have higher $\mathscr{P}_c$, less power is allocated in order to limit the interference inflicted on the primary user transmission. Also, the proposed power control policy  depends on sensing performance through $\mathscr{P}_c$, which is a function of detection and false alarm probabilities, $P_{\nid}$ and $P_{\f}$, respectively.
\end{Rem}

\begin{Rem}
By setting $\alpha=0$ in (\ref{eq:EE_concave}) in Appendix \ref{appendix2}, the optimization problem becomes
\begin{align}
\max_{
\substack{P(g,h) \geq 0}} R_{\avg}
\end{align}
which corresponds to the throughput maximization problem. Therefore, solving the above optimization problem or equivalently inserting $\alpha=0$ into the proposed scheme in (\ref{eq:opt_power}), we can readily obtain the optimal power control strategy that maximizes the average throughput of secondary users in the presence of unslotted primary users.
\end{Rem}

\begin{Rem}
By inserting $\alpha=0$, $\mathscr{P}_{c,0}=0$ and $\mathscr{P}_{c,1}=1$ into (\ref{eq:opt_power}), we can see that the optimal power control scheme has a similar structure to the scheme that maximizes the throughput of secondary users operating over a single frequency band given in \cite[eq. (36)]{stotas}, where it is assumed that the primary users do not change their activity during the entire frame duration of the secondary users, i.e., a time-slotted transmission scheme. Hence, our results can be specialized to the time-slotted case by setting $\mathscr{P}_{c,0}$ and $\mathscr{P}_{c,1}$ equal to $0$ and $1$, respectively.
\end{Rem}
\subsection{Peak Transmit Power and Average Interference Power Constraints}
In this subsection, we consider that the secondary user transmission is subject to peak transmit power and average interference power constraints. Under these assumptions, the optimization problem can be expressed as
\begin{align}
\label{eq:EE_max_Ppk_Qavg}
&\hspace{-1.6cm}\max_{
\substack{T_{\f}, P(g,h)}} \eta_{EE}=\frac{R_{\avg}}{\big(\frac{T_{\f}-\tau}{T_{\f}}\big)\Pr\{\hH_0\}\E_{g,h}\{P(g,h)\}+P_{c_r}} \\ \label{eq:coll_constraint2}
\text{subject to }&\mathscr{P}_c \leq \mathscr{P}_{c,max} \\  \label{eq:peak_transmit_power}
&P(g,h) \le P_{\pk} \\  \label{eq:avg_inter_power1}
&\Big(\frac{T_{\f}-\tau}{T_{\f}}\Big)\mathscr{P}_c\Pr\{\hH_0\}\E_{g,h}\big\{P(g,h)|g|^2\big\} \le Q_{\avg} \\
& P(g,h)  \geq 0 \\ \label{eq:T_lower}
& T_{\f}  \geq \tau,
\end{align}
where $P_{\pk}$ represents the peak transmit power limit at the secondary transmitter. Subsequently, the optimal power control policy is determined in the following result.

\begin{Lem1}\label{teo:2} For a given frame duration $T_{\f}$, the optimal power control scheme subject to the constraints in (\ref{eq:peak_transmit_power}) -- (\ref{eq:T_lower}) is obtained as
\begin{align} \label{eq:opt_power_Ppk}
&\hspace{1cm}P_{\text{opt}}(g,h) =\min\bigg\{\bigg[\frac{A_1+\sqrt{\Delta_1}}{2}\bigg]^{+},P_{\pk} \bigg\} \\
\text{where } &A_1=\frac{\log_2(e)}{\alpha+\mu\mathscr{P}_c|g|^2}-\frac{2N_0+\sigma_s^2}{|h|^2} \\
&\hspace{-1cm}\Delta_1\!=\!A_1^2\!-\!\frac{4}{|h|^2}\bigg(\!\frac{N_0(N_0\!+\!\sigma_s^2)}{|h|^2}\!-\frac{\log_2(e)(N_0+(1\!-\!\mathscr{P}_c) \sigma_s^2)}{\alpha+\mu\mathscr{P}_c|g|^2}\!\bigg).
\end{align}
Above, $\mu$ is the Lagrange multiplier associated with the average interference power constraint in (\ref{eq:avg_inter_power1}).
\end{Lem1}
Since we follow similar steps as in the proof of Theorem \ref{teo:1}, the proof is omitted for the sake of brevity.
\begin{Rem}
Different from the optimal power control strategy in Theorem \ref{teo:1}, the instantaneous transmission power level in (\ref{eq:opt_power_Ppk}) is limited by $P_{\pk}$ due to the peak transmit power constraint, which imposes stricter limitations than the average transmit power constraint.
\end{Rem}

\begin{Rem}
Setting $\alpha=0$ in (\ref{eq:opt_power_Ppk}), we obtain the optimal power control strategy which maximizes the throughput of secondary users with unslotted primary users, which is in agreement with the result derived in \cite{ozcan}.
\end{Rem}

Algorithm 1 can be easily modified to maximize the EE of the secondary users under peak transmit power and average interference constraints by calculating the power level, $P_{\text{opt}}(g,h)$ through the expression in (\ref{eq:opt_power_Ppk}) and updating the Lagrange multiplier $\mu$ similarly as in (\ref{eq:subgradient_Qavg}).

\section{Spectrally-Efficient Optimal Power Control and Frame Duration with a Minimum EE Constraint} \label{sec:SE_power}
\subsection{Average Transmit Power and Average Interference Power Constraints}
In this subsection, we analyze the EE-SE tradeoff by formulating the optimal power control problem to maximize the average throughput of the secondary users subject to a minimum EE constraint, and average transmit power, average interference power and collision constraints. The optimization problem is formulated as follows:
\begin{align}
\label{eq:throughput_max_Pavg_Qavg}
&\hspace{0.8cm}\max_{
\substack{T_{\f} \geq \tau,P(g,h) \geq 0}} R_{\avg}\\  \label{eq:coll_constraint3}
\text{subject to } &\mathscr{P}_c \leq \mathscr{P}_{c,max} \\
\label{eq:EEmin_Pavg_Qavg} &\frac{R_{\avg}}{\big(\frac{T_{\f}-\tau}{T_{\f}}\big)\Pr\{\hH_0\}\E_{g,h}\{P(g,h)\}+P_{c_r}} \geq \text{EE}_{\text{min}}\\ \label{eq:avg_inter_power3}
&\Big(\frac{T_{\f}-\tau}{T_{\f}}\Big)\mathscr{P}_c\Pr\{\hH_0\}\E_{g,h}\big\{P(g,h)|g|^2\big\} \le Q_{\avg}  \\
\label{eq:avg_transmit_power3}
&\Big(\frac{T_{\f}-\tau}{T_{\f}}\Big)\Pr\{\hH_0\}\E_{g,h}\big\{P(g,h)\big\} \le P_{\avg},
\end{align}
where $\text{EE}_{\text{min}}$ denotes the minimum required EE. The optimal power control is determined in two steps. In the first step, we determine the average power level at which the required minimum EE is achieved. In the second step, we optimally allocate the transmission power in order to maximize the average throughput of the secondary users by combining the power level obtained in the first step under average transmit power and average interference power constraints. In this regard, we first provide the following result.
\begin{Prop} \label{prop:2} For a given frame duration $T_{\f}$, the average power level that satisfies the minimum required EE can be obtained as
\begin{align} \label{eq:Pavg_EEmin}
P_{\avg}^*=\Big(\frac{T_{\f}-\tau}{T_{\f}}\Big)\Pr\{\hH_0\}\E_{g,h}\big\{P^*(g,h)\big\}
\end{align}
and $P^*(g,h)$ is given by
\begin{align}  \label{eq:P0_EEmin}
&\hspace{2.4cm}P^*(g,h) =\bigg[\frac{A_3+\sqrt{\Delta_3}}{2}\bigg]^{+},\\ \label{eq:A3}
\text{where } &A_3=\frac{(1+\eta)\log_2(e)}{\eta \text{EE}_{\text{min}}}-\frac{2N_0+\sigma_s^2}{|h|^2} \\ \nonumber
&\hspace{-0.6cm}\Delta_3\!=\!A_3^2\!-\!\frac{4}{|h|^2}\bigg(\!\frac{N_0(N_0\!+\!\sigma_s^2)}{|h|^2}\! \\&\label{eq:delta3}-\frac{(1\!+\!\eta)\log_2(e)(N_0\!+\!(1\!-\!\Pr\{\hH_0\}\mathscr{P}_c) \sigma_s^2)}{\eta \text{EE}_{\text{min}}}\!\bigg).
\end{align}
The optimal value of $\eta$ can be found by solving the equation below:
\begin{align} \label{eq:eta_EEmin}
R_{\avg}+ \eta \bigg(\Big(\frac{T_{\f}-\tau}{T_{\f}}\Big)\Pr\{\hH_0\}\E\big\{P^*(g,h)\big\}\bigg)=0.
\end{align}
\end{Prop}
\emph{Proof:} See Appendix \ref{appendix3}.

Using the results in Proposition \ref{prop:2}, the throughput optimization problem subject to the minimum EE constraint is equivalent to the throughput maximization under an average power constraint with the power limit, $P^*_{\avg}$, which achieves the minimum required EE. By combining this power limit with the average transmit power constraint in (\ref{eq:avg_transmit_power3}), we define the operating average transmission power as follows:
\begin{align} \label{eq:Pavg_optimal_minEE}
P_{op}=\begin{cases} P^*_{\avg} &\text{if } P_{\avg} \geq P^*_{\avg} \\ P_{\avg} &\text{if } P_{\avg} < P^*_{\avg} \\& \text{and} \left. \eta_{EE} \right |_{\text{s.t.} (\ref{eq:avg_transmit_power})  \text{ and } (\ref{eq:avg_inter_power})} \geq \text{EE}_{\text{min}} \\
0 &\text{if } P_{\avg} < P^*_{\avg} \\& \text{and} \left. \eta_{EE} \right |_{\text{s.t.} (\ref{eq:avg_transmit_power}) \text{ and } (\ref{eq:avg_inter_power})} < \text{EE}_{\text{min}}
\end{cases}
\end{align}

\begin{figure*}[htb]
\centering
\includegraphics[width=1\textwidth]{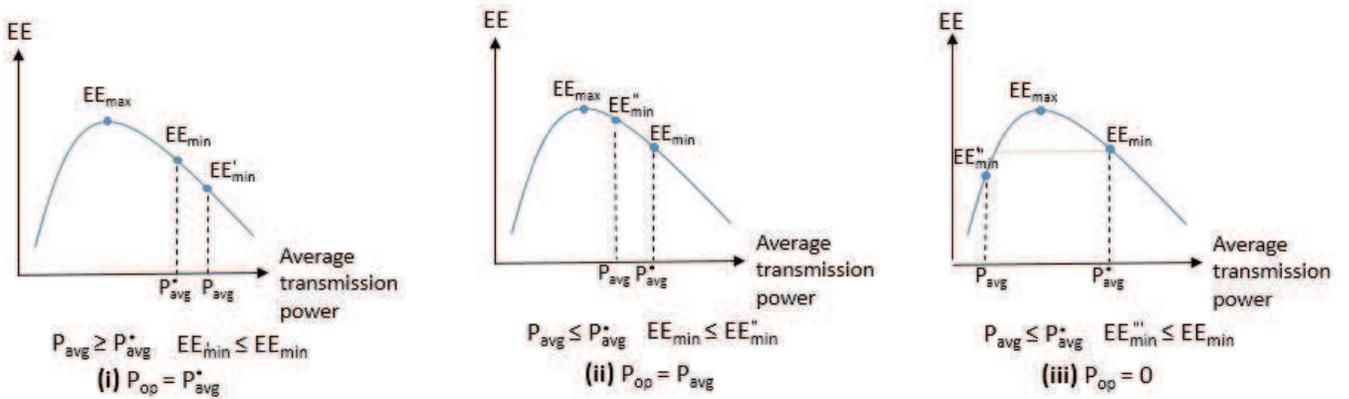}
\caption{The operating average transmission power for three cases.}
\label{fig:Op_power}
\end{figure*}

The operating average transmission power is determined according to three cases as illustrated in Fig. \ref{fig:Op_power}. \\
{\bf \emph{Case (i)}}: When $P_{\avg}$ is larger than $P^*_{\avg}$, average transmit power constraint $P_{\avg}$ is loose. Since operating at average transmission power level greater than $P^*_{\avg}$ violates the minimum required EE constraint, we set $P_{op}=P^*_{\avg}$ and the optimal transmission power control policy is obtained by satisfying $P^*_{\avg}$ with equality. This case is illustrated in Fig. \ref{fig:Op_power}.(i). \\
{\bf \emph{Case (ii)}}: As shown in Fig. \ref{fig:Op_power}.(ii), when $P_{\avg}$ is less than $P^*_{\avg}$ and the EE achieved at $P_{\avg}$ is greater than $\text{EE}_{\text{min}}$, average transmit power constraint $P_{\avg}$ is dominant.  Since average transmission power is limited by $P_{\avg}$, we set $P_{op}=P_{\avg}$ and the optimal transmission power control policy is found when $P_{\avg}$ is satisfied with equality.  \\
{\bf \emph{Case (iii)}}: As demonstrated in Fig. \ref{fig:Op_power}.(iii), when $P_{\avg} < P^*_{\avg}$ and the EE achieved at $P_{\avg}$ is less than $\text{EE}_{\text{min}}$, there is no feasible solution, and hence we set $P_{op}=0$

In the following result, we identify the optimal power control strategy.
\begin{Lem1}\label{teo:3} For a given frame duration $T_{\f}$, if  $P_{\avg} < P^*_{\avg}$ and the maximum EE subject to the constraints in (\ref{eq:avg_transmit_power}) and (\ref{eq:avg_inter_power}) is less than $ \text{EE}_{\text{min}}$, the power level is set to zero, i.e., $P^*_0(g,h) =0$, otherwise we allocate the power according to
\begin{align} \label{eq:opt_power4}
&\hspace{2cm}P_{\text{opt}}(g,h) =\bigg[\frac{A_4+\sqrt{\Delta_4}}{2}\bigg]^{+}
\end{align}

\begin{align} \label{eq:A4}
\text{where } &A_4=\frac{\log_2(e)}{\vartheta+\varphi\mathscr{P}_c|g|^2}-\frac{2N_0+\sigma_s^2}{|h|^2} \\
\label{eq:delta4}
&\hspace{-0.9cm}\Delta_4\!=\!A_4^2\!-\!\frac{4}{|h|^2}\bigg(\!\frac{N_0(N_0\!+\!\sigma_s^2)}{|h|^2}\!-\frac{\log_2(e)(N_0\!+\!(1\!-\!\mathscr{P}_c) \sigma_s^2)}{\vartheta+\varphi\mathscr{P}_c|g|^2}\!\bigg).
\end{align}
Above, $\vartheta$ and $\varphi$ are the Lagrange multipliers associated with the average transmit power constraint, $\min(P_{\avg},P^*_{\avg})$ and interference power constraint in (\ref{eq:avg_inter_power3}), respectively.
\end{Lem1}
\emph{Proof:} See Appendix \ref{appendix4}.

In Table \ref{table:algorithm2}, we provide the details of an algorithm for jointly finding the optimal power control policy and frame duration that maximize the average throughput of the secondary users subject to constraints on collision duration ratio, the minimum required EE, average transmit power and interference power in the presence of unslotted primary users.

\begin{table}
\caption{} \label{table:algorithm2}
\begin{algorithm}[H]
    \caption{The optimal power control and frame duration algorithm that maximizes the average throughput of the secondary users under the minimum EE, average transmit power, average interference power, and collision constraints}
    \begin{algorithmic}[1]
      \STATE For a given $P_{\nid}$, $P_{\f}$, $\mathscr{P}_{c,max}$, $\text{EE}_{\text{min}}$, initialize $\eta^{(0)}=\eta_{\text{init}}$,  $\vartheta^{(0)}=\vartheta_{\text{init}}$, $\varphi^{(0)}=\varphi_{\text{init}}$
     \IF{$\mathscr{P}_{c,max} < \Pr\{\mH_1|\hH_0\}$}
      \STATE $T_{\f,\text{opt}}=0, P_{\text{opt}}(g,h)=0$
     \ELSE
\STATE Find the optimal value of $\eta$ that solves (\ref{eq:eta_EEmin}) by using a root-finding function.
\STATE Calculate $P_{\avg}^*=\Big(\frac{T_{\f}-\tau}{T_{\f}}\Big)P(\hH_0)\E_{g,h}\big\{P^*(g,h)\big\}$ where $P^*(g,h)$ is given in (\ref{eq:P0_EEmin}).
\IF{$P_{\avg} < P^*_{\avg}  \text{ and } \left. \eta_{EE} \right |_{\text{s.t.} (\ref{eq:avg_transmit_power}) \text{ and } (\ref{eq:avg_inter_power})} < \text{EE}_{\text{min}}$}
\STATE $P_{\text{opt}}(g,h)=0$
\ELSE
\STATE $P_{op}=\min(P_{\avg},P^*_{\avg})$ and calculate $P_{\text{opt}}(g,h)$ using (\ref{eq:opt_power4})
\STATE Update $\vartheta$ and $\varphi$ using subgradient method
 \ENDIF
\STATE Calculate $R_{\avg}$ using (\ref{eq:Ravg})
\STATE $T_{\f,opt}=\argmax R_{\avg}$ by bisection search
\STATE $P^*_{\text{opt}}(g,h)=[P_{\text{opt}}(g,h)]_{T_{\f}=T_{\f,opt}}$
 \ENDIF
    \end{algorithmic}
  \end{algorithm}
\vspace{-0.9cm}
\end{table}

\subsection{Peak Transmit Power and Average Interference Power Constraints}
In this subsection, we consider that the objective function in (\ref{eq:throughput_max_Pavg_Qavg}) is subject to the constraints in (\ref{eq:coll_constraint3})- (\ref{eq:avg_inter_power3}) and the peak transmit power constraint $P(g,h) < P_{pk}$ instead of the average transmit power constraint. In this case, we derive the optimal power control as follows:

\begin{Lem1}\label{teo:4}The average power level at which the minimum required EE is achieved can be determined by inserting the power control given below in (\ref{eq:P0_EEmin_Ppk}) into (\ref{eq:Pavg_EEmin}):
\begin{align}  \label{eq:P0_EEmin_Ppk}
P^*(g,h) =\bigg\{\bigg[\frac{A_3+\sqrt{\Delta_3}}{2}\bigg]^{+},P_{\pk}\bigg\},
\end{align}
where $A_3$ and $\Delta_3$ are given in (\ref{eq:A3}) and (\ref{eq:delta3}), respectively. If the maximum EE at $P_{\pk}$ is less than $ \text{EE}_{\text{min}}$, the power level is set to zero, i.e., $P_{\text{opt}}(g,h) =0$, otherwise the optimal power control can be found as
\begin{align} \label{eq:opt_power5}
P_{\text{opt}}(g,h) =\min \bigg\{ \bigg[\frac{A_4+\sqrt{\Delta_4}}{2}\bigg]^{+},P_{\pk} \bigg\}
\end{align}
Above, $A_4$ and $\Delta_4$ are given in (\ref{eq:A4}) and (\ref{eq:delta4}), respectively.
\end{Lem1}
\emph{Proof:} We follow similar steps as in the proof of Proposition \ref{prop:2} and Theorem \ref{teo:3} with peak transmit power constraint in consideration. Therefore, the power levels are limited by $P_{\pk}$ in this case. \hfill $\square$

\section{Numerical Results}\label{sec:num_results}
In this section, we present and discuss the numerical results for the optimal power control and frame duration, which maximize the EE or throughput of the secondary users, and analyze the resulting collisions with the unslotted primary users. Unless mentioned explicitly, the noise variance is $N_0=0.01$ and the variance of primary user's received signal is $\sigma_s^2=0.1$. Also, the mean values of the durations of ON and OFF periods, denoted by $\lambda_0$ and $\lambda_1$, are set to $650$ ms and $352$ ms, respectively so that $\Pr\{\mH_0\} \approx 0.65$, corresponding to the setting in the voice over Internet protocol (VoIP) traffic. The step size $t$ and tolerance $\epsilon$ are chosen as  $0.1$ and $10^{-5}$, respectively. The circuit power $P_{c_r}$ is set to $1$. We consider a Rayleigh fading environment, and hence the channel power gains of the transmission link and interference link are exponentially distributed with unit mean.

It is assumed that the secondary users employ energy detection scheme for spectrum sensing, and hence the probabilities of detection and false alarm are expressed, respectively as \cite{tandra}, \cite{liang}
\begin{align}
P_{\nid}&=\mathcal{Q}\bigg(\Big(\frac{\varepsilon}{N_0}-\frac{\sigma_s^2}{N_0}-1\Big)\sqrt{\frac{\tau f_s}{2\frac{\sigma_s^2}{N_0}+1}}\bigg)\\
P_{\f}&=\mathcal{Q}\bigg(\Big(\frac{\varepsilon}{N_0}-1\Big)\sqrt{\tau f_s}\bigg),
\end{align}
where $\mathcal{Q}(x)=\int_{x}^{\infty}\frac{1}{\sqrt{2\pi}}\rme^{-t^2/2}dt$ is the Gaussian $Q$-function, $\varepsilon$ represents the decision threshold and $f_s$ denotes the sampling frequency. The decision threshold $\varepsilon$ can be chosen to satisfy the target detection and false alarm probabilities, denoted by $\bar{P}_{\nid}$ and $\bar{P}_{\f}$, respectively and the resulting sensing duration $\tau$ is expressed as
\begin{align}
\tau=\frac{1}{f_s}\bigg(\frac{\mathcal{Q}^{-1}(\bar{P}_{\f})-\sqrt{2\sigma_s^2+1}\mathcal{Q}^{-1}(\bar{P}_{\nid})}{\sigma_s^2}\bigg)^2.
\end{align}
In the numerical computations, $f_s$ is set to $100$ kHz.

\begin{figure}[ht]
\centering
{\includegraphics[width=0.5\textwidth]{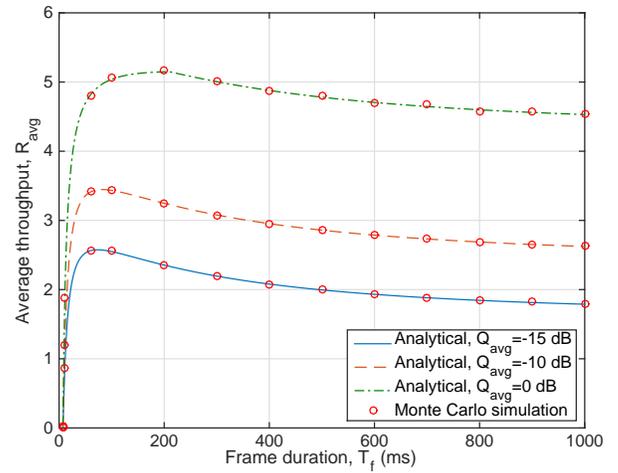}}
\caption{Average throughput of the secondary users, $R_{\avg}$ vs. frame duration, $T_{\f}$.}\label{fig:R_Tf}
\end{figure}

\begin{figure*}[ht!]
\centering
\begin{subfigure}[b]{0.32\textwidth}
\centering
\includegraphics[width=\textwidth]{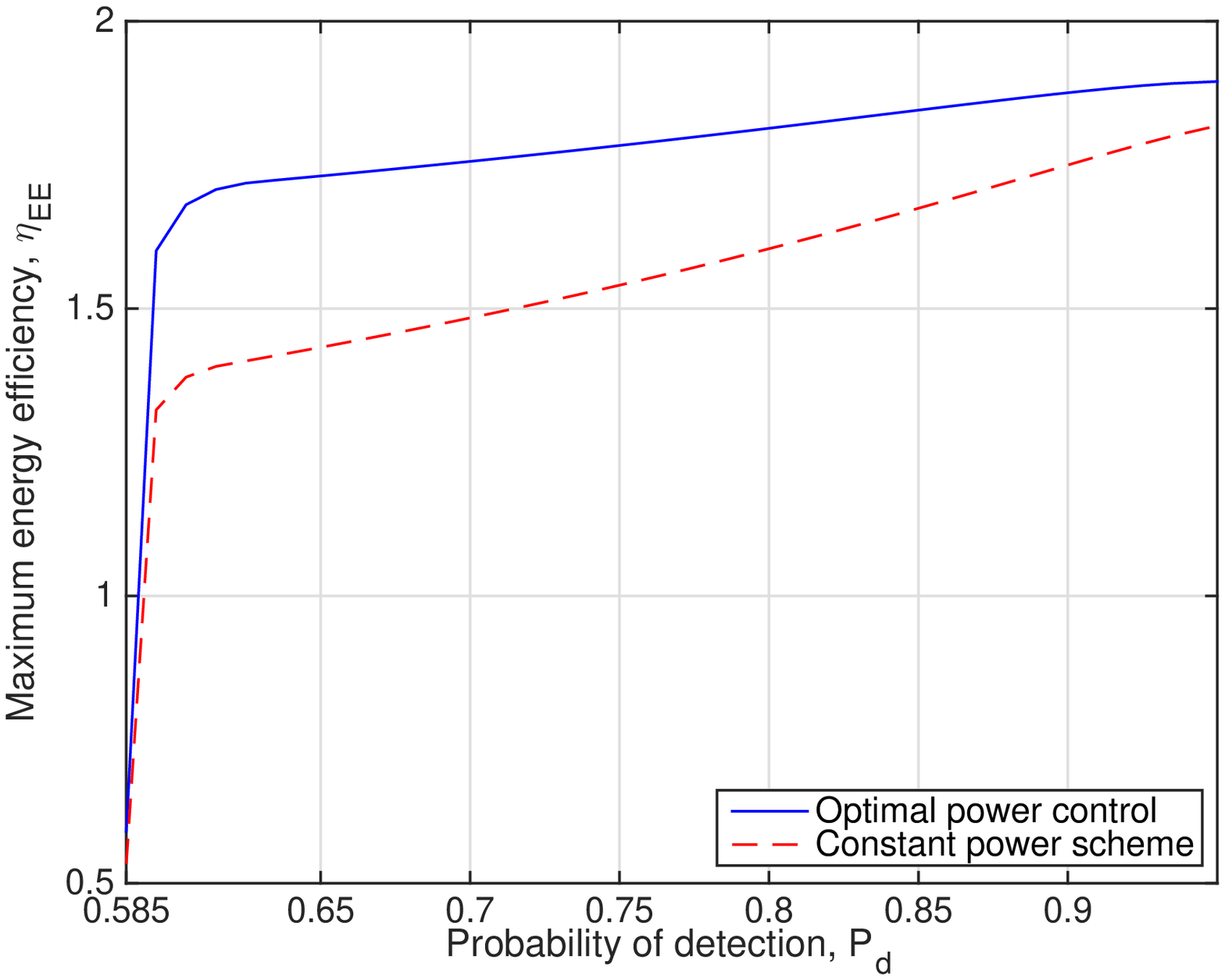}
\caption{}
\end{subfigure}
\begin{subfigure}[b]{0.32\textwidth}
\centering
\includegraphics[width=\textwidth]{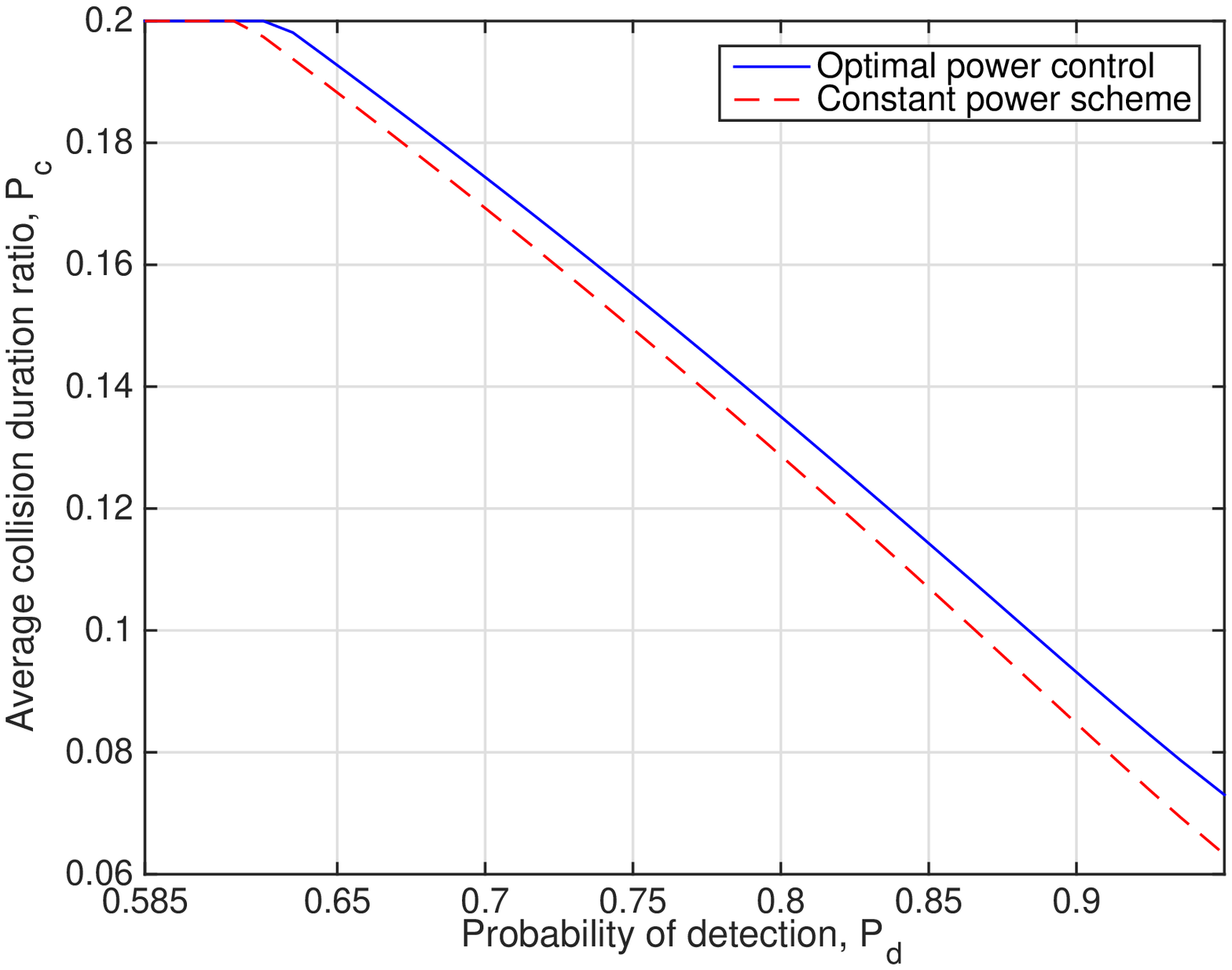}
\caption{}
\end{subfigure}
\begin{subfigure}[b]{0.32\textwidth}
\centering
\includegraphics[width=\textwidth]{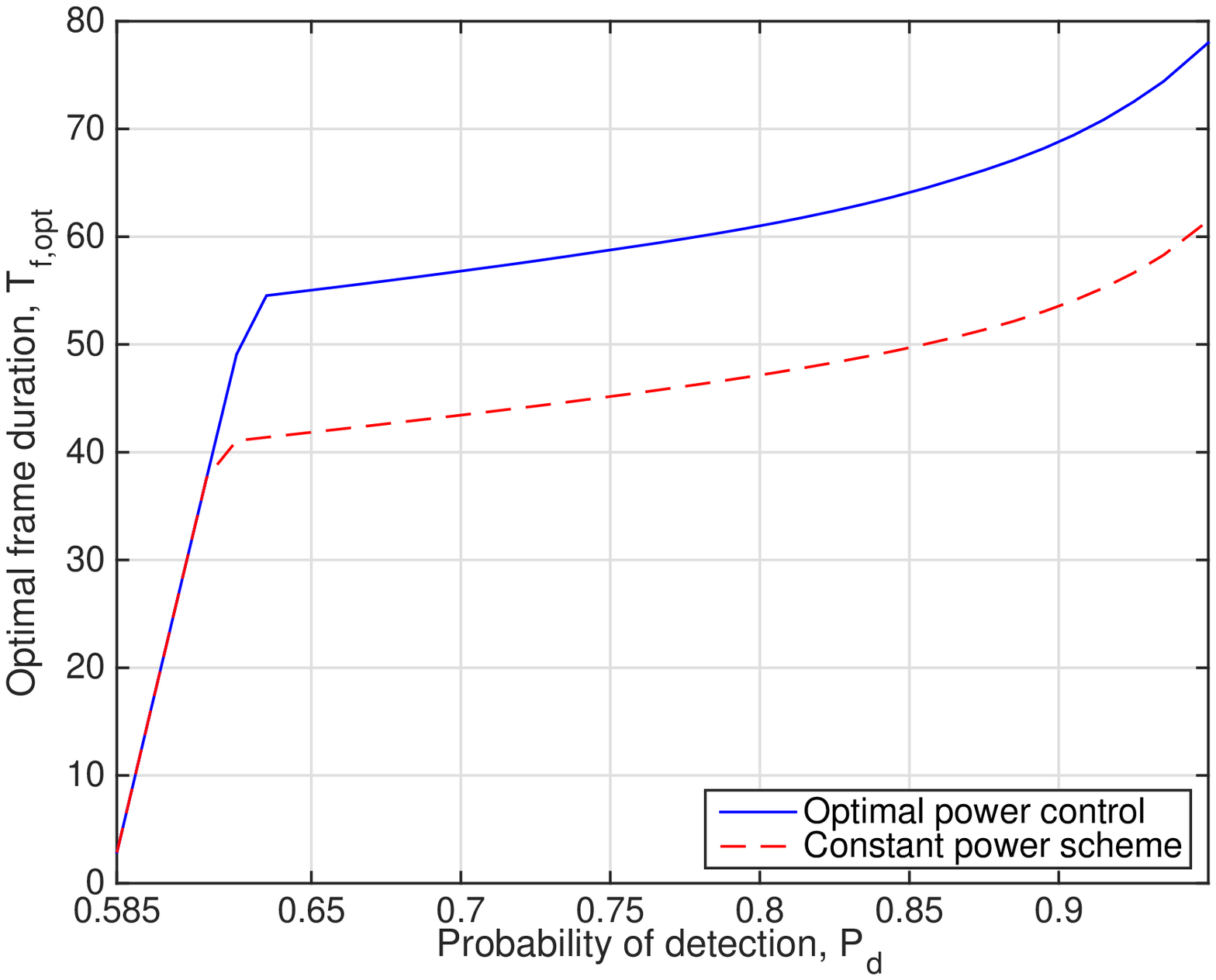}
\caption{}
\end{subfigure}
\caption{\small{(a) Maximum EE of the secondary users vs. the probability of detection, $P_{\nid}$ (b) Average collision duration ratio, $\mathscr{P}_c$ vs. $P_{\nid}$ (c) Optimal frame duration, $T_{\f,opt}$ vs. $P_{\nid}$.}}\label{fig:EE_Topt_Pd}
\end{figure*}

\begin{figure*}[h]
\centering
\begin{subfigure}[b]{0.32\textwidth}
\centering
\includegraphics[width=\textwidth]{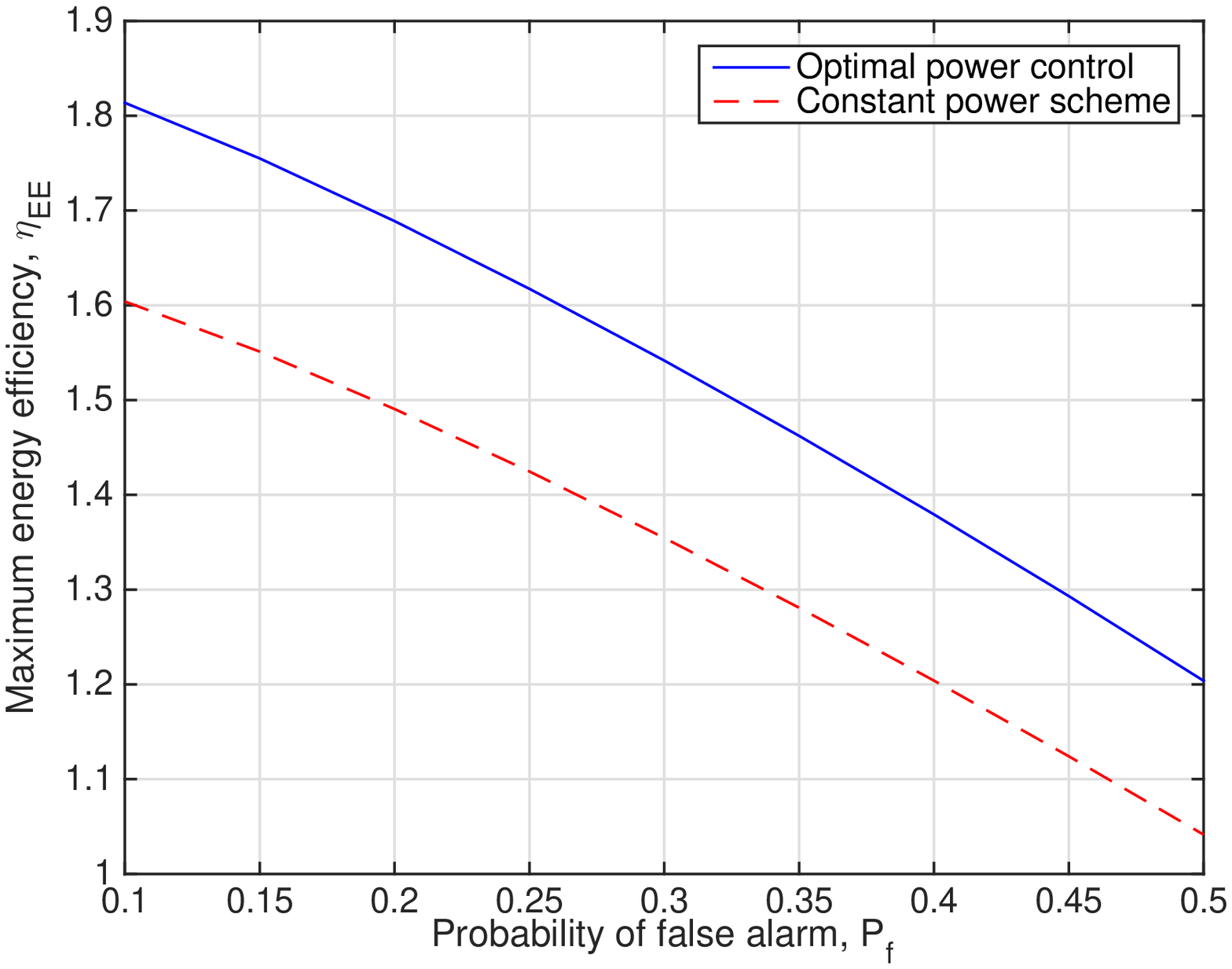}
\caption{}
\end{subfigure}
\begin{subfigure}[b]{0.32\textwidth}
\centering
\includegraphics[width=\textwidth]{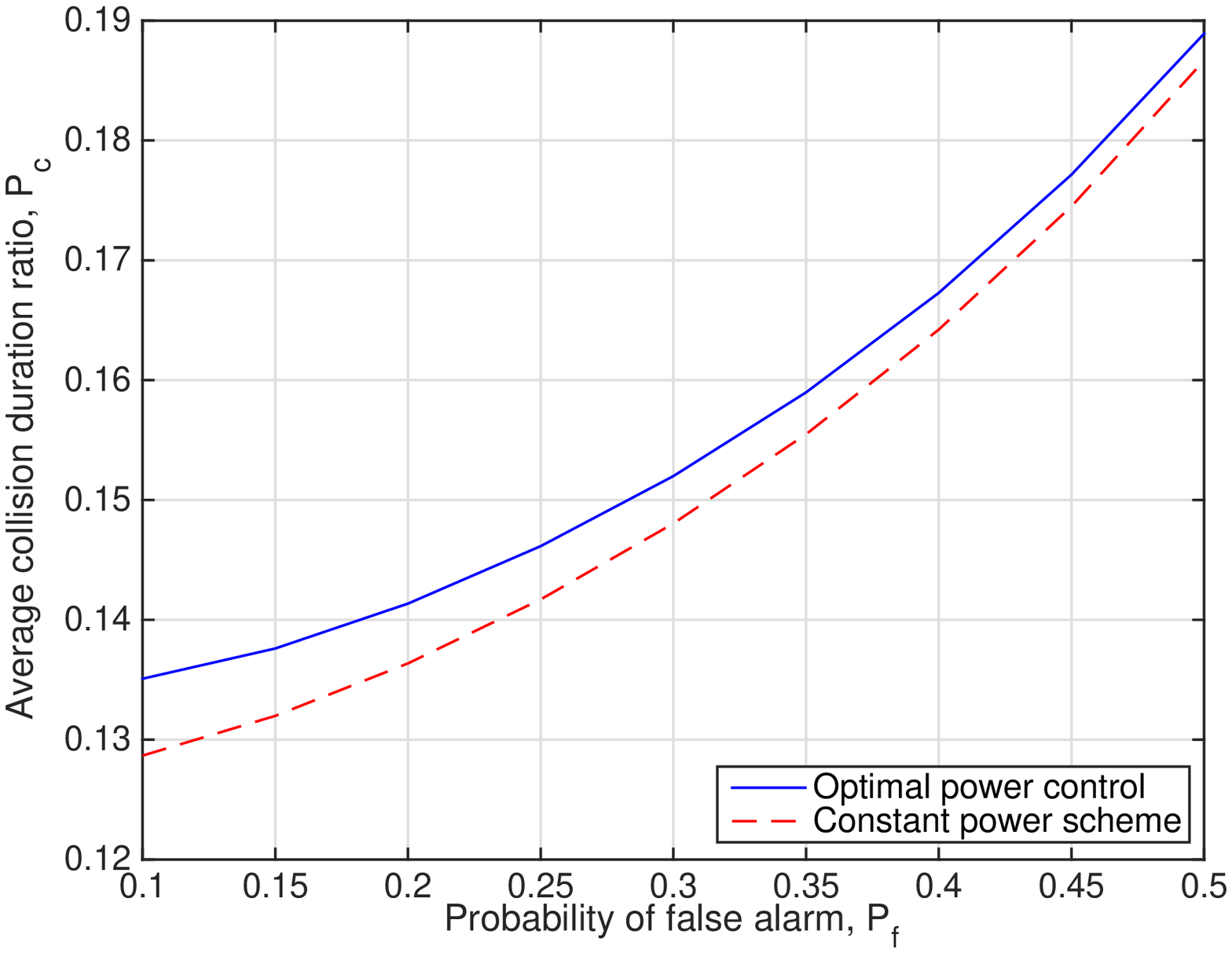}
\caption{}
\end{subfigure}
\begin{subfigure}[b]{0.32\textwidth}
\centering
\includegraphics[width=\textwidth]{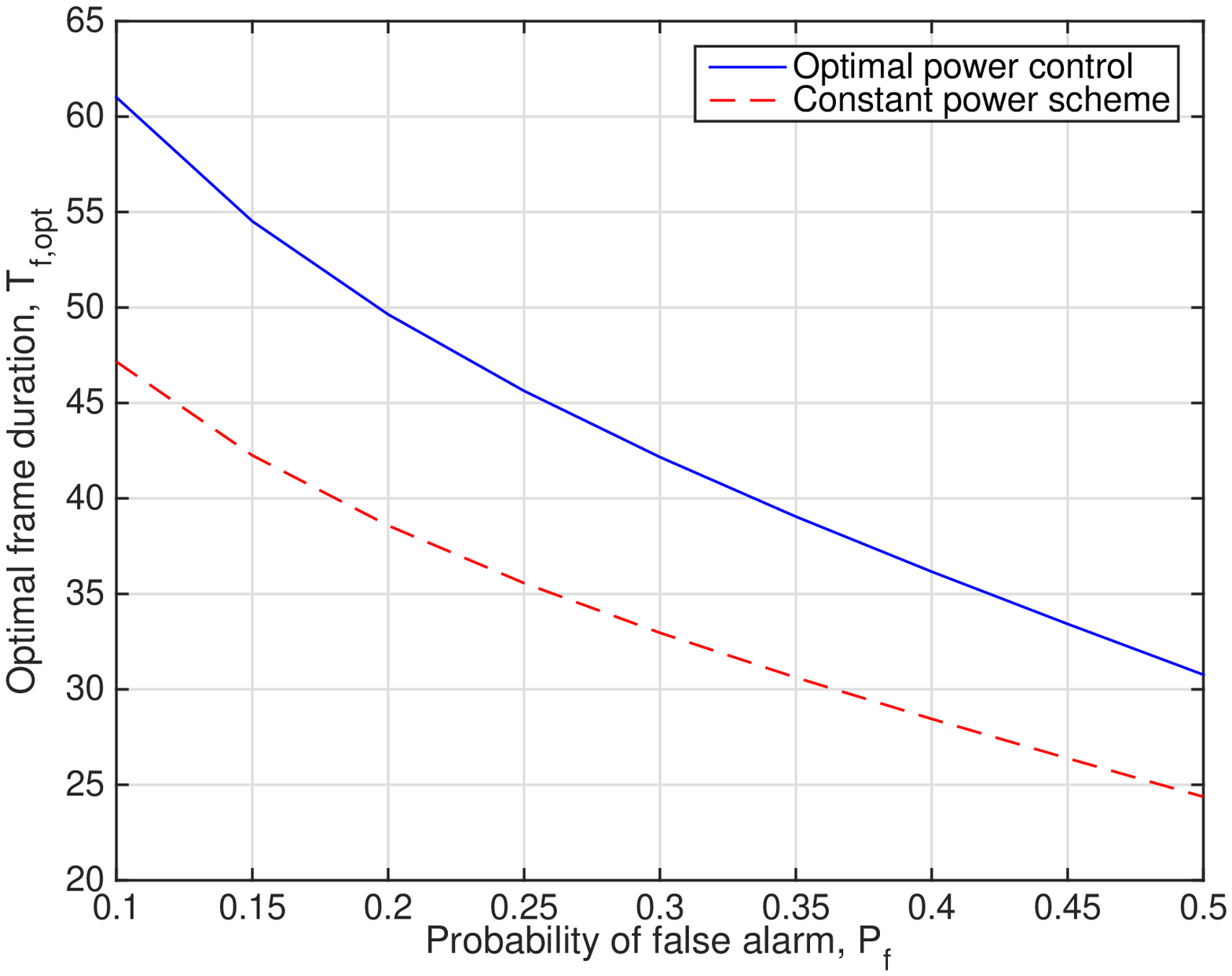}
\caption{}
\end{subfigure}
\caption{\small{(a) Maximum EE of the secondary users vs. the probability of false alarm, $P_{\f}$ (b) Average collision duration ratio, $\mathscr{P}_c$ vs. $P_{\f}$ (c) Optimal frame duration, $T_{\f,opt}$ vs. $P_{\f}$.}}\label{fig:EE_Topt_Pf}
\end{figure*}

In Fig.~\ref{fig:R_Tf}, we plot the average throughput of the secondary users, $R_{\avg}$, as a function of the frame duration $T_{\f}$ for $P_{\pk}=10$ dB and  different average power constraints, namely $Q_{\avg}=-15$ dB, $Q_{\avg}=-10$ dB and $Q_{\avg}=0$ dB. We consider target detection probability $\bar{P}_{\nid}=0.9$ and false alarm probability $\bar{P}_{\f}=0.1$, and hence $\tau$ becomes $7.21$ ms. Transmission power level is chosen according to $\min\Big\{P_{\pk},\Big(\frac{T_{\f}}{T_{\f}-\tau}\Big)\frac{Q_{\avg}}{\Pr\{\mH_0,\hH_0\}\mathscr{P}_{c,0}+\Pr\{\mH_1,\hH_0\}\mathscr{P}_{c,1}}\Big\}$. In this setting, average throughput formulation in (\ref{eq:Ravg}) is also verified through Monte Carlo simulations with $100000$ runs. It is seen that $R_{\avg}$ initially increases with increasing transmission duration. After reaching a peak value, $R_{\avg}$ begins to diminish as the secondary user starts colliding with primary user transmissions more frequently, degrading the performance. It is also observed that as the interference power constraint gets looser, i.e., as $Q_{\avg}$ changes from $-15$ to $0$ dB, higher throughput is achieved since secondary user transmits at higher power levels. As illustrated in the figure, $R_{\avg}$ is not a concave function of $T_{\f}$. However, $R_{\avg}$ curves are seen to exhibit a quasiconcave property and there exists an optimal frame duration that maximizes the throughput.

In Fig.~\ref{fig:EE_Topt_Pd}, we display the maximum EE $\eta_{EE}$, average collision duration ratio $\mathscr{P}_c$, and the optimal frame duration $T_{\f,opt}$ as functions of the probability of detection $P_{\nid}$. We set the maximum collision limit as $\mathscr{P}_{c,max}=0.2$. It is assumed that the average transmit power constraint is $P_{\avg}=10$ dB and average interference power constraint is $Q_{\avg}=-20$ dB. We consider both the transmission with the optimal power control policy and constant-power transmission. For the constant power case, power is not adaptively varied with respect to the channel power gains of the transmission link and interference link. On the other hand, optimal power control derived in (\ref{eq:opt_power}) is a function of both $h$ and $g$. As $\bar{P}_{\nid}$ increases while keeping $\bar{P}_{\f}$ fixed at $0.1$ and hence sensing performance improves, secondary user has a higher EE. In addition, collision duration ratio decreases with increasing detection probability in both cases of optimal power control and constant power. For $P_{\nid}$ values less than $0.585$, collision constraint is not satisfied for any value of the frame duration $T_{\f}$, and therefore the secondary user throughput is $0$. When $P_{\nid}$ takes values between $0.585$ and $0.6$, maximum EE is achieved at the maximum collision limit, i.e, when $\mathscr{P}_c=0.2$. It is also observed that the optimal power control leads transmissons with a larger frame duration while satisfying the maximum allowed collision limit and achieving a higher EE compared to constant-power transmissions.

In Fig.~\ref{fig:EE_Topt_Pf}, we plot the maximum EE $\eta_{EE}$, average collision duration ratio $\mathscr{P}_c$, and the optimal frame duration $T_{\f,opt}$ as functions of the probability of false alarm $P_{\f}$. We consider the same setting as in the previous figure. It is seen that as $P_{\f}$ increases while keeping $P_{\nid}$ fixed at $0.9$, sensing performance degrades and secondary users experience more false alarm events, which leads to more collisions with the primary user transmission. Therefore, secondary user has a lower EE in both cases of optimal power control and constant power. We also notice in Fig.~\ref{fig:EE_Topt_Pf}(a) that the optimal power control outperforms constant-power transmissions.

\begin{figure}[ht]
\centering
{\includegraphics[width=0.5\textwidth]{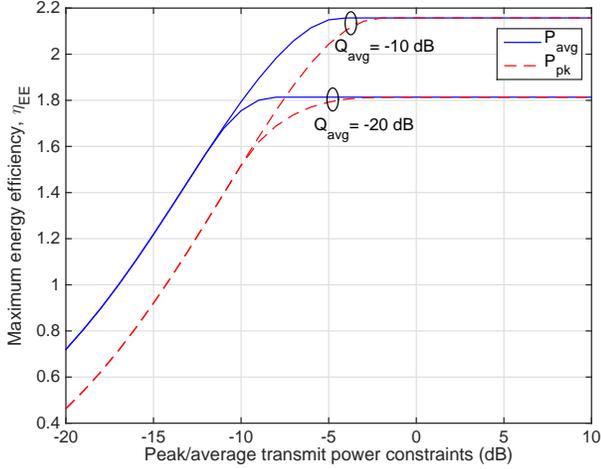}}
\vspace{-0.4cm}
\caption{Maximum EE of the secondary users, $\eta_{EE}$ vs. peak/average transmit power constraints.}\label{fig:EE_Pavg_Ppk}
\end{figure}

The maximum EE, $\eta_{EE}$, as a function of peak/average transmit power constraints is illustrated in Fig.~\ref{fig:EE_Pavg_Ppk}. Regarding the average interference constraint, we consider two scenarios: $Q_{\avg}=-10$ dB and $Q_{\avg}=-20$ dB. Target probabilities of detection and false alarm are set to $0.8$ and $0.1$, respectively, for which the corresponding sensing duration is $4.85$ ms. In addition, the frame duration is selected to maximize the EE. It can be seen from the figure that for low values of $P_{\avg}$ and $P_{\pk}$, average interference power constraints are loose, and hence the power is determined by either the average or peak transmit power constraint, which results in the same EE regardless of whether $Q_{\avg}=-10$ dB or $Q_{\avg}=-20$ dB. The EE of the secondary users increases with increasing peak/average transmit power levels. As expected, peak transmit power constraint yields lower EE compared to that achieved under the average transmit power constraint since the instantaneous transmission power is limited by $P_{\pk}$ under the peak transmit power constraint, which imposes stricter limitations than the average transmit power constraint. As the constraints become less stringent and the peak and average transmit power levels are further increased, the maximum EE levels off and becomes the same under peak/average transmit power constraints since the power starts being allocated according to only the average interference constraint, $Q_{\avg}$, due to this constraint being the dominant one.
\begin{figure}[ht]
\centering
{\includegraphics[width=0.5\textwidth]{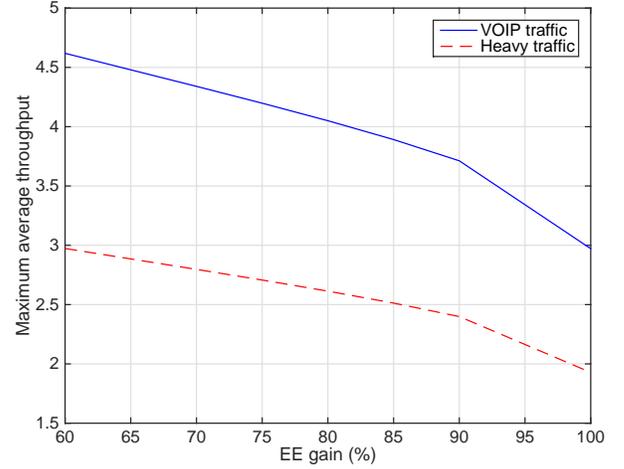}}
\vspace{-0.4cm}
\caption{Maximum average throughput vs. EE gain.}\label{fig:throughput_EEmin_gain}
\end{figure}

In Fig.~\ref{fig:throughput_EEmin_gain}, we display the maximum average throughput as a function of the EE gain in percentage for different levels of primary traffic. More specifically, we consider a normal traffic load, i.e., VOIP traffic with $\lambda_ 0 = 650$ ms and $\lambda_1 = 352$ ms as assumed before, and also heavy traffic load with $\lambda_0=350$ ms and $\lambda_1=650$ ms so that $\Pr\{\mH_0\} \approx 0.37$. It is assumed that $\mathscr{P}_{c,max}=0.3$, average transmit power constraint is $P_{\avg}=0$ dB and average interference power constraint is $Q_{\avg}=10$ dB, and $\bar{P}_{\nid}=0.8$, $\bar{P}_{\f}=0.1$, and hence $\tau=4.85$ ms. The frame duration for normal traffic and heavy traffic are chosen optimally as $T_{\f}=125$ ms and $T_{\f}=36$ ms, respectively, in order to maximize the EE in each traffic model. The EE gain is calculated as the ratio of the minimum required EE, $\text{EE}_{\text{min}}$, to the maximum EE achieved with the proposed power control in (\ref{eq:opt_power}). It is seen that a tradeoff between the EE and SE indeed exists, i.e., as the EE gain increases, the maximum average throughput of the secondary users decreases. We also note that the primary user with a heavy traffic load occupies the channel more often, and hence the secondary users have less opportunity to access the channel. In this heavy-load scenario, secondary users experience more frequent collisions with the primary user transmission. As a result, secondary users have lower throughput in the presence of heavy primary-user traffic compared to the case with a normal primary-user traffic.

\begin{figure}[ht]
\centering
{\includegraphics[width=0.5\textwidth]{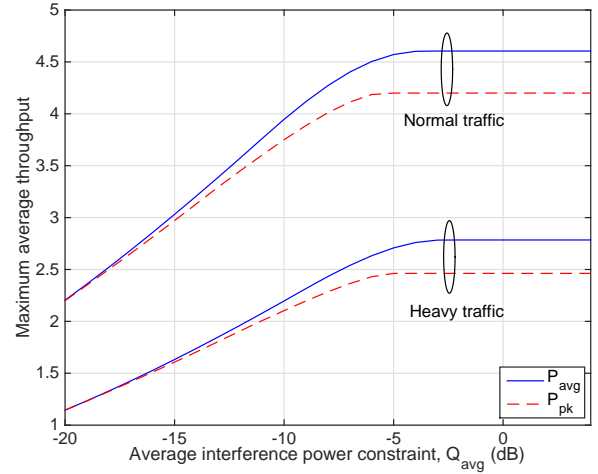}}
\vspace{-0.4cm}
\caption{Maximum average throughput vs. average interference power constraint, $Q_{\avg}$ under a minimum EE constraint.}\label{fig:throughput_EEmin_Qavg}
\end{figure}
In Fig.~\ref{fig:throughput_EEmin_Qavg}, we display the maximum average throughput as a function of the average interference power constraint, $Q_{\avg}$, under a minimum EE constraint, namely $\text{EE}_{\min}=1$ bit/joule in the presence of primary users with normal and heavy traffic loads. The frame duration is selected to maximize the system performance for each case. We assume imperfect spectrum sensing with $P_{\nid}=0.8$ and set $P_{\f}=0.1$ and $P_{\avg}=P_{\pk}=4$ dB, $\mathscr{P}_{c,max}=0.3$. As $Q_{\avg}$ increases, the secondary users transmit with higher power levels, resulting in higher throughput. However, increasing $Q_{\avg}$ further than a certain threshold does not provide performance improvements since the power starts being limited by either $P_{\avg}$ or $P_{\pk}$. In addition, secondary users have higher throughput with longer transmission duration when the primary user has a normal traffic load rather than a heavy one.

\section{Conclusion}\label{sec:conclusion}
In this paper, we have derived the optimal power control policies that maximize the EE or maximize the average throughput of the secondary users while satisfying a minimum required EE level, in the presence of unslotted primary users, imperfect sensing, and average/peak transmit power, average interference power and collision constraints. We have also provided low-complexity algorithms to jointly optimize the transmission power and frame duration. Numerical results reveal important relations and tradeoffs between the EE and throughput performance of the secondary users. We have addressed how secondary user's EE, collisions with the primary user transmissions, and the optimal frame duration vary as a function of the probabilities of detection and false alarm. It is also shown that optimal power control policy significantly enhances the system performance compared to the constant power scheme. The impact of the primary traffic on the system performance is analyzed as well. In particular, we have observed that secondary users achieve smaller throughput when the primary user has a heavy traffic load.
\appendix
\subsection{Proof of Proposition \ref{prop:1}} \label{appendix1}
\emph{Proof:} The first derivative of $\mathscr{P}_c$ with respect to frame duration $T_{\f}$ is given (\ref{eq:Pc-derivative}) at the top of next page.
\begin{figure*}
\begin{align}
\small
\begin{split}
\frac{\partial \mathscr{P}_c}{\partial T_{\f}}&=\Big(\Pr\{\mH_0|\hH_0\}\lambda_0\Pr\{\mH_1\}^2-\Pr\{\mH_1|\hH_0\}\lambda_1\Pr\{\mH_0\}^2\Big) \bigg(\frac{1-\rme^{-\frac{T_{\f}-\tau}{\lambda_0 \Pr\{\mH_1\}}}}{(T_{\f}-\tau)^2}-\frac{1}{\lambda_0\Pr\{\mH_1\}(T_{\f}-\tau)} \rme^{-\frac{T_{\f}-\tau}{\lambda_0 \Pr\{\mH_1\}}}\bigg).
\end{split}\label{eq:Pc-derivative}
\normalsize
\end{align}
\hrule
\end{figure*}
The expression inside the first parenthesis can easily be seen to be greater than zero if $P_{\f}<P_{\nid}$ and less than zero if $P_{\f}>P_{\nid}$ by using the formulations in (\ref{eq:prior_probs}), (\ref{eq:PH0_hH0_exp}) and (\ref{eq:PH1_hH0_exp}). In order to show that the expression inside the second parenthesis is always nonnegative, we compare it with zero as follows:
\begin{align}\label{eq:second_paranthesis}
\frac{1-\rme^{-\frac{T_{\f}-\tau}{\lambda_0 \Pr\{\mH_1\}}}}{(T_{\f}-\tau)^2}-\frac{1}{\lambda_0\Pr\{\mH_1\}(T_{\f}-\tau)} \rme^{-\frac{T_{\f}-\tau}{\lambda_0 \Pr\{\mH_1\}}} > 0.
\end{align}
Above inequality can be rewritten as
\begin{align}\label{eq:comparision}
\bigg(1+\frac{T_{\f}-\tau}{\lambda_0\Pr\{\mH_1\}}\bigg)\rme^{-\frac{T_{\f}-\tau}{\lambda_0 \Pr\{\mH_1\}}}<1.
\end{align}
Left-hand side of (\ref{eq:comparision}) is a decreasing function since its first derivative with respect to frame duration $T_{\f}$ is $-\frac{T_{\f}-\tau}{(\lambda_0 \Pr\{\mH_1\})^2}\rme^{-\frac{T_{\f}-\tau}{\lambda_0 \Pr\{\mH_1\}}} \le 0$. Since it is a decreasing function and it takes values between $(0,1)$ for $T_{\f}> \tau$, the inequality in (\ref{eq:comparision}) and hence the inequality in (\ref{eq:second_paranthesis}) hold.
With this, we have shown that the expression inside the second parenthesis in (\ref{eq:Pc-derivative}) is nonnegative, and therefore the first derivative of $\mathscr{P}_c$ is greater than zero if $P_{\f}<P_{\nid}$ and less than zero if $P_{\f}>P_{\nid}$, proving the property that $\mathscr{P}_c$ is increasing with $T_{\f}$ if $P_{\f}<P_{\nid}$ and decreasing with $T_{\f}$ if $P_{\f}>P_{\nid}$.

Also, it can be easily verified that $\mathscr{P}_c$ takes values between $\Pr\{\mH_1|\hH_0\}$ and $\Pr\{\mH_1\}$. In particular, we examine the limit of $\mathscr{P}_c$ as $T_{\f}$ approaches $\tau$ and $\infty$ as follows:
\begin{align} \nonumber
\lim_{T_{\f} \rightarrow \tau} \mathscr{P}_c & =  \Pr\{\mH_0|\hH_0\}\bigg(\Pr\{\mH_1\}-\frac{\lambda_0 \Pr\{\mH_1\}^2}{\lambda_0\Pr\{\mH_1\}}\bigg) \\ \nonumber & \hspace{0.8cm} + \Pr\{\mH_1|\hH_0\}\bigg(\Pr\{\mH_1\}+\frac{\lambda_1 \Pr\{\mH_0\}^2}{\lambda_0\Pr\{\mH_1\}}\bigg) \\ & = \Pr\{\mH_1|\hH_0\} \\ \nonumber
\lim_{T_{\f} \rightarrow \infty} \mathscr{P}_c & =  \Pr\{\mH_0|\hH_0\}\Pr\{\mH_1\}+ \Pr\{\mH_1|\hH_0\}\Pr\{\mH_1\}  \\& =\Pr\{\mH_1\}
\end{align}

\subsection{Proof of Theorem \ref{teo:1}} \label{appendix2}
\emph{Proof:} The objective function in (\ref{eq:EE_max_Pavg_Qavg}) is quasiconcave since the average throughput in the numerator is composed of positive weighted sum of logarithms which are strictly concave and the power consumption in the denominator is both affine and positive. Therefore, the optimal power value can be found iterativaly by using Dinkelbach's method \cite{dinkelbach}. The optimization problem is first transformed into the equivalent parameterized concave problem as follows:
\begin{align}
\label{eq:EE_concave}
&\hspace{-2cm}\max_{
\substack{P(g,h) \geq 0}} \bigg\{R_{\avg}-\alpha \bigg(\Big(\frac{T_{\f}-\tau}{T_{\f}}\Big)\Pr\{\hH_0\}\E\{P(g,h)\}+P_{c_r}\bigg)\bigg\} \\ \label{eq:avg_transmit_power_eq}
\text{subject to } &\Big(\frac{T_{\f}-\tau}{T_{\f}}\Big)\Pr\{\hH_0\}\E\big\{P(g,h)\big\} \le P_{\avg} \\  \label{eq:avg_inter_power_eq}
&\Big(\frac{T_{\f}-\tau}{T_{\f}}\Big)\mathscr{P}_c\Pr\{\hH_0\}\E\big\{P(g,h)|g|^2\big\} \le Q_{\avg},
\end{align}
where $\alpha$ is a nonnegative parameter. At the optimal value of $\alpha^*$, the following condition is satisfied
\begin{align} \label{eq:F_alpha}
F(\alpha^*)\!=\!R_{\avg}\!-\!\alpha^*\bigg(\!\Big(\frac{T_{\f}-\tau}{T_{\f}}\Big)\Pr\{\hH_0\}\E\{P(g,h)\}\!+\!P_{c_r}\!\bigg)\!=\!0.
\end{align}
Explicitly, the solution of $F(\alpha^*)$ is equivalent to the solution of the EE maximization problem in (\ref{eq:EE_max_Pavg_Qavg}). It is shown that Dinkelbach's method converges to the optimal solution at a superlinear convergence rate. The detailed proof of convergence and further details can be found in \cite{schaible}. Since the parameterized problem in (\ref{eq:EE_concave}) is concave for a given $\alpha$, the optimal power levels can be obtained by using the Lagrangian optimization approach as follows:
\begin{align}
\small
\begin{split}  \label{eq:lagrangian_function}
&\mathcal{L}(P(g,h),\lambda,\nu,\alpha)\!=\!R_{\avg}-\alpha \bigg(\!\Big(\frac{T_{\f}\!-\!\tau}{T_{\f}}\Big)\Pr\{\hH_0\}\E\{P(g,h)\}\!+\!P_{c_r}\!\!\bigg) \\ &\hspace{1.8cm}-\lambda\bigg(\Big(\frac{T_{\f}-\tau}{T_{\f}}\Big)\Pr\{\mH_0\}\E\{P(g,h)\}-P_{\avg}\bigg) \\
&\hspace{1.8cm}-\nu\bigg(\Big(\frac{T_{\f}-\tau}{T_{\f}}\Big)\mathscr{P}_c\Pr\{\hH_0\}\E\big\{P(g,h)|g|^2\big\}-Q_{\avg} \bigg),
\end{split}
\normalsize
\end{align}
where $\lambda$  and $\nu$ are the nonnegative Lagrange multipliers. The Lagrange dual problem is defined as
\begin{align}
\min_{\lambda, \nu \ge 0} \max_{P(g,h) \geq 0} \mathcal{L}(P(g,h),\lambda,\nu,\alpha).
\end{align}
For fixed $\lambda$ and $\nu$ values, and each fading state, we express the subproblem using the Lagrange dual decomposition method \cite{palomar}. According to the Karush-Kuhn-Tucker (KKT) conditions, the optimal power control $P_{\text{opt}}(g,h)$ must satisfy the set of equations and inequalities below:
\begin{align} \nonumber
& \frac{\Pr\{\hH_0\}}{\log_e(2)}\Big(\frac{T_{\f}-\tau}{T_{\f}}\Big) \bigg[\bigg(\frac{(1-\mathscr{P}_c)|h|^2}{N_0+P_{\text{opt}}(g,h)|h|^2}\bigg)\\ \nonumber &\!+\! \bigg(\frac{\mathscr{P}_c|h|^2}{N_0\!+\!\sigma_s^2+P_{\text{opt}}(g,h)|h|^2}\bigg)\bigg]\! -\! (\alpha\!+\!\lambda)\Big(\frac{T_{\f}-\tau}{T_{\f}}\Big) \Pr\{\hH_0\} \\ \label{eq:KKT_condition1} &\hspace{2.8cm}-\nu\Big(\frac{T_{\f}-\tau}{T_{\f}}\Big)  \Pr\{\hH_0\}\mathscr{P}_c|g|^2=0\\ \label{eq:average_transmit_cond3} &\lambda\bigg(\Big(\frac{T_{\f}-\tau}{T_{\f}}\Big)\Pr\{\hH_0\}\E\{P_{\text{opt}}(g,h)\}-P_{\avg}\bigg)=0,
\\ \label{eq:average_inter_cond3}&\nu\bigg(\Big(\frac{T_{\f}-\tau}{T_{\f}}\Big)\mathscr{P}_c\Pr\{\hH_0\}\E\big\{P_{\text{opt}}(g,h)|g|^2\big\}-Q_{\avg} \bigg) = 0, \\ \label{eq:lagrange_multipliers}
&\lambda \ge 0, \nu \ge 0.
\end{align}

Solving (\ref{eq:KKT_condition1}) and incorporating the nonnegativity of the transmit power yield the desired result in (\ref{eq:opt_power}). \hfill $\square$
\subsection{Proof of Proposition \ref{prop:2}} \label{appendix3}
In order to find the operating power level, which satisfies the minimum required EE, we consider that the objective function in (\ref{eq:throughput_max_Pavg_Qavg}) is subject to only a minimum EE constraint in (\ref{eq:EEmin_Pavg_Qavg}). Since $R_{\avg}$ is a concave function of the transmission power and the feasible set defined by the minimum EE constraint is a convex set, KKT conditions are both sufficient and necessary for the optimal solution. The constraint in (\ref{eq:EEmin_Pavg_Qavg}) can be rewritten as follows
\begin{align}
\hspace{-0.2cm}R_{\avg}- \text{EE}_{\text{min}} \Big(\big(\frac{T_{\f}-\tau}{T_{\f}}\big)\Pr\{\hH_0\}\E\{P(g,h)\}+P_{c_r}\Big) \geq 0.
\end{align}
By defining $\eta$ as the Lagrange multiplier associated with the above constraint, the Lagrangian function is expressed as
\begin{align} \nonumber
&\mathcal{L}(P(g,h),\eta)=(1+\eta)R_{\avg} \\&\hspace{1cm}-\eta \text{EE}_{\text{min}} \Big(\big(\frac{T_{\f}-\tau}{T_{\f}}\big)\Pr\{\hH_0\}\E\{P(g,h)\}+P_{c_r}\Big).
\end{align}
By setting the derivative of the above function with respect to $P(g,h)$ equal to zero at the optimal power level, we obtain the equation in (\ref{eq:Lagrangian_EEmin}) given at the top of next page.
\begin{figure*}
\begin{align}
\begin{split} \label{eq:Lagrangian_EEmin}
\hspace{-0.4cm}&\left. \frac{\partial \mathcal{L}(P(g,h),\eta)}{\partial P(g,h)} \right |_{P(g,h)=P^*(g,h)} \!\!\!\!\!=(1+\eta)\frac{\Pr\{\hH_0\}}{\log_e(2)}\Big(\frac{T_{\f}\!-\!\tau}{T_{\f}}\Big)\bigg[\bigg(\frac{(1-\mathscr{P}_c)|h|^2}{N_0+P^*(g,h)|h|^2}\bigg)\!+\!\bigg(\frac{\mathscr{P}_c|h|^2}{N_0+\sigma_s^2+P^*(g,h)|h|^2}\bigg)\bigg]\\&\hspace{13cm}-\eta \text{EE}_{\text{min}}\Big(\frac{T_{\f}-\tau}{T_{\f}}\Big)  \Pr\{\hH_0\}\!=\!0.
\end{split}
\end{align}
\hrule
\end{figure*}
\normalsize
Solving the equation in (\ref{eq:Lagrangian_EEmin}) leads to the desired characterization in (\ref{eq:P0_EEmin}) and the Lagrange multiplier, $\eta$ can be determined by satisfying the minimum EE constraint with equality or solving (\ref{eq:eta_EEmin}). Consequently, the average transmission power is obtained by inserting (\ref{eq:P0_EEmin}) into (\ref{eq:Pavg_EEmin}). \hfill $\square$
\subsection{Proof of Theorem \ref{teo:3}} \label{appendix4}
The Lagrangian function is expressed as
\begin{equation}
\small
\begin{split}
&\mathcal{L}(P(g,h),\vartheta,\varphi)=R_{\avg} - \vartheta \Big(\Big(\frac{T_{\f}-\tau}{T_{\f}}\Big)P(\hH_0)\E\{P(g,h)\} \\&\!-\!\min(P_{\avg},\!P^*_{\avg})\!\Big)\!-\!\varphi \Big(\!\Big(\frac{T_{\f}\!-\!\tau}{T_{\f}}\Big)\mathscr{P}_c\Pr\{\hH_0\}\E\{\!P(g,\!h)|g|^2\!\}\!-\!Q_{\avg})\!\Big).
\end{split}
\normalsize
\end{equation}
Setting the derivative of the above function with respect to transmission power, $P(g,h)$, to zero and arranging the terms give the desired optimal power control in (\ref{eq:opt_power4}).\hfill $\square$


\end{document}